\documentclass[aps,twocolumn,showpacs,amsmath,amssymb]{revtex4-1}
\usepackage{amsmath}
\usepackage{amssymb}
\usepackage[colorlinks,plainpages=false,linkcolor=black,urlcolor=blue,citecolor=blue,pdfpagemode=UseNone,pdfstartview=FitH]{hyperref}
\usepackage[T1]{fontenc}
\usepackage[utf8]{inputenc}
\usepackage[english]{babel}
\usepackage[left=2.5cm,right=2.5cm,top=2cm,bottom=3.0cm,a4paper]{geometry}
\usepackage{color}
\usepackage{float}
\usepackage{graphicx}
\usepackage{booktabs}
\usepackage{tabularx}
\usepackage{mathtools}

\newcommand{\CaK}{CaKFe$_4$As$_4$}
\newcommand{\Ca}{CaFe$_2$As$_2$}
\newcommand{\K}{KFe$_2$As$_2$}

\newcommand{\bea}{\begin{eqnarray}}
\newcommand{\eea}{\end{eqnarray}}
\newcommand{\beq}{\begin{equation}}
\newcommand{\eeq}{\end{equation}}
\newcommand{\bd}{\begin{displaymath}}
\newcommand{\ed}{\end{displaymath}}
\newcommand{\ba}{\begin{array}}
	\newcommand{\ea}{\end{array}}
\newcommand{\bi}{\begin{itemize}}
	\newcommand{\ei}{\end{itemize}}
\newcommand{\bc}{\begin{center}}
	\newcommand{\ec}{\end{center}}
\newcommand{\bfl}{\begin{flushleft}}
	\newcommand{\efl}{\end{flushleft}}
\newcommand{\bfr}{\begin{flushright}}
	\newcommand{\efr}{\end{flushright}}

\def\6{\partial}

\def\={\!\!\!&=&\!\!\!}
\def\+{\!\!\!&&\!\!\!+~}
\def\-{\!\!\!&&\!\!\!-~}

\newcolumntype{C}{@{}>{\hspace*{5mm}}c}

\newcommand{\sig}{\sigma}
\newcommand{\sigb}{\bar{\sigma}}
\newcommand{\sigs}{\sigma^\prime}

\newcommand{\RN}[1]{\text{\uppercase\expandafter{\romannumeral #1\relax}}}

\newcommand{\be}{\begin{equation}}
\newcommand{\ee}{\end{equation}}

\def\6{\partial}

\def\={\!\!\!&=&\!\!\!}
\def\+{\!\!\!&&\!\!\!+~}
\def\-{\!\!\!&&\!\!\!-~}

\begin{document}

\title{Electronic properties, low-energy Hamiltonian and superconducting instabilities in \CaK\ }

\author{Felix Lochner$^{1,2}$, Felix Ahn$^{2}$, Tilmann Hickel$^{1}$, and Ilya Eremin$^{2}$}

\affiliation{$^1$Max-Planck-Institut f\"ur Eisenforschung, D-40237 D\"usseldorf, Germany}

\affiliation{$^2$Institut f\"ur Theoretische Physic III, Ruhr-University Bochum, D-44801 Bochum, Germany}

\begin{abstract}
	We analyze the electronic properties of the recently discovered stoichiometric
	superconductor \CaK\ by combining an \textit{ab initio} approach and a projection of the band structure to a low-energy tight-binding Hamiltonian, based on the maximally localized Wannier orbitals of the 3$d$ Fe states. We identify the key symmetries as well as differences and similarities in the electronic structure between \CaK\ and the parent systems \Ca\ and \K. In particular, we find \CaK\ to have a significantly more quasi-two-dimensional electronic structure than the latter systems. Finally, we study the superconducting instabilities in \CaK\ by employing the leading angular harmonics approximation (LAHA) and find two potential A$_{1g}$-symmetry representation of the superconducting gap to be the dominant instabilities in this system. 
\end{abstract}

\date{\today}


\maketitle

\section{Introduction}
The discovery of the iron-based superconductors (FeSC) in 2008 created a wide and highly interesting field in solid state physics \cite{Kamihara.JACS.130.3296(2008)}.
Most of the FeSCs are magnetic metals for the stoichiometric composition and superconduct once the magnetism is destroyed by pressure, disorder or doping, which results in complex
phase diagrams with rich physics \cite{Johnston_review,Hirschfeld_review,Chubukov_review}. The highest $T_c$
is often found at fractional compositions, which due to disorder represents a
serious challenge in understanding the pairing mechanism in these systems. In particular, for the non-phononic mechanisms of Cooper-pairing,  with anisotropic or sign-changing superconducting gaps, any disorder adds extra
complications due to a necessity to quantify the pair-breaking effects. Thus, the presence of the few stoichiometric FeSCs offers a unique opportunity to study the phenomenon of superconductivity in these compounds with much higher accuracy both experimentally and theoretically. Among them the recently discovered
\CaK\ (CaK1144) shows a particularly
high value of the superconducting transition temperature $T_c \approx 35.8$\,K and the upper critical field $H^c_{c2} \approx 71$\,T \cite{Iyo.JACS.138.3410(2016),Mou.PRL.117.277001(2016),Yang.PRB.95.064506(2017),Cho.PRB.95.100502(2017),Shi.JPSJ.85.124714(2016),Meier2016}.
At the same time if one considers the equal ratio
of Ca and K in \CaK\ one could compare this system with hole-doped Ba$_{0.46}$K$_{0.54}$Fe$_2$As$_2$, which
has a similar $T_c$ of 34\,K \cite{Liu.PRB.89.134504(2014),Cho2016} but is randomly disordered on the single (Ba/K) site. 

\begin{figure}[h]
	\centering
	\includegraphics[width=7.6cm]{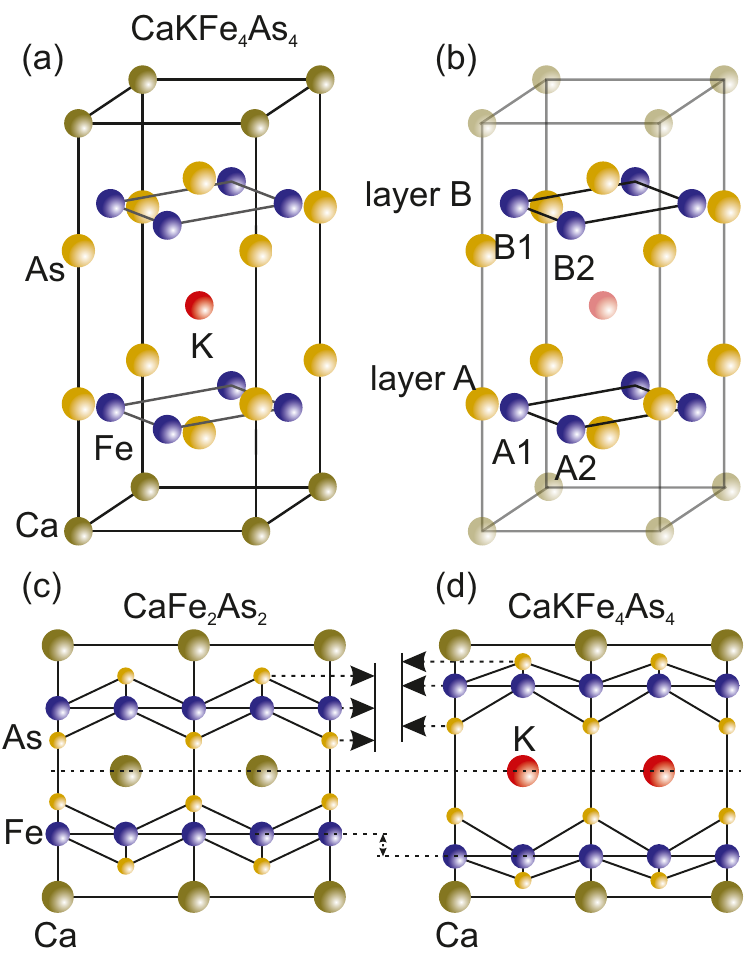}
	\caption{(color online) Atomic structure of \CaK\ (a) and the definition of the two different FeAs-layers (layer A/B) (b). In (c) and (d) the space symmetries of the atom positions of \Ca\ and \CaK\ are illustrated. For \Ca\ the Fe-atoms are in high-symmetry points at $(0,\frac{1}{2},\pm\frac{1}{4})$ and $(\frac{1}{2},0,\pm\frac{1}{4})$. The distance of the As-atoms above and below the Fe-atoms is equivalent. For \CaK\ the FeAs-layers are shifted away from the high-symmetry points, also the distance of the As-atoms is different for the upper and the lower case.}
	\label{Fig:Sketch_atomPos_1144_122}
\end{figure}

Overall \CaK\ is one of the most important representatives of the \textit{AeA}1144 structure family, consisting of alkaline-earth
(\textit{Ae}) and alkali
 (\textit{A}) metals. This structure modifies the intensively studied 122 materials such that the atom in the middle of the unit cell is replaced by an alkali 
metal atom (see Fig..~\ref{Fig:Sketch_atomPos_1144_122}a). This substitution changes the space group from $I4/mmm$ to the non-symmetric group $P4/mmm$, since the different \textit{Ae} and \textit{A} layers cause a shift of the intermediate FeAs-layer  out of their high-symmetry positions. Moreover, in those materials the off-positions of the As-atoms lead to two different Fe-As distances.

To identify the impact on the superconducting performance, it is decisive to investigate the interplay between crystal and electronic structure in particular stoichiometric iron-based superconductors. To achieve this goal, we investigate the electronic structure of \CaK\ with \textit{ab-initio} methods using density functional theory (DFT). In particular, we systematically compare the electronic structure of the CaK1144 and the \K \ and \Ca  (122) materials and analyze doping as well as the influences of the off-symmetry positions of the FeAs-layer in \CaK. In addition we develop the low-energy description of this system by employing a tight-binding (TB) parametrization of the electronic structure using maximally localized $3d$~Wannier orbitals and discuss the different symmetries within the system. This Hamiltonian will then be used to analyze the superconducting gap function by using the leading angular harmonic approximation (LAHA) method. \cite{Maiti2011a,Maiti2011b,Ahn.PRB.89.144513(2014),Maiti.PRB.85.014511(2012)}

The paper is organized as follows. In Section \ref{sec1]} we present the
results of our DFT calculations and compare the electronic properties of CaK1144 and 122 structures. In Section \ref{sec2} we perform the tight-binding parametrization and discuss the symmetries of the low-energy Hamiltonian. Section \ref{sec3} presents the analysis of the superconducting gap symmetries using LAHA and comparison with the state of art ARPES data. Finally, we conclude the results of our study in Section \ref{sec4}.

\section{Electronic structure \label{sec1]}}
For the DFT part of our work, we use the Vienna \textit{ab initio} simulation package (VASP) \cite{Kresse.PRB.47.558(1993),Kresse.ComMatSci.6.15(1996),Kresse.PRB.54.11169(1996)} with the projector augmented wave (PAW) \cite{Blochl.PRB.50.17953(1994)} basis and employ experimentally obtained crystal parameters for \CaK, \Ca\ and \K\ as measured previously \cite{Iyo.JACS.138.3410(2016),Rozsa.ZNB.36.1668(1981),Kreyssig.PRB.78.184517(2008)}. Hereby we use VASP in the generalized gradient approximation (GGA) \cite{Perdew.PRL.77.3865(1996)}. For a better convergence we also consider the $p$-orbital and $s$-orbital states of the Fe- and As-ions as valence states.

For the tight-binding calculations we use the wannier90 package \cite{Mostofi.CompPhysCom.185.2309(2014)} with the in VASP implemented vasp2wannier interface, which calculates the maximally localized Wannier functions (MLWF) \cite{Marzari.PRB.56.12847(1997)}. Here we focus only on the Fe $3d$-orbitals and neglect the $p$-orbitals. Their influence we will discuss later.\\

\begin{figure}[h]
	\centering
	\includegraphics[width=7.6cm]{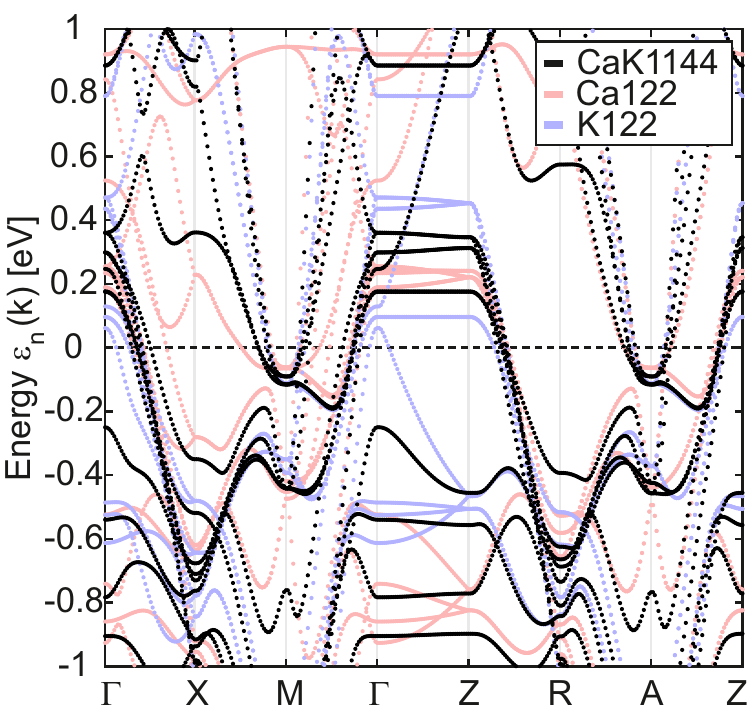}
	\caption{(color online) Electronic dispersion for \CaK\ (black), \Ca\ (shaded red) and \K\ (shaded blue) along the high-symmetry directions of the first Brillouin 
	zone obtained from DFT using experimentally determined lattice parameters \cite{Iyo.JACS.138.3410(2016),Kreyssig.PRB.78.184517(2008),Rozsa.ZNB.36.1668(1981)}. For better comparison, the electronic structures 
	of \Ca\ and \K\  were shifted with a rigid band shift to acquire the same doping level as \CaK .  }
	\label{Fig:Bands_all_Doped}
\end{figure}

In Fig.~\ref{Fig:Bands_all_Doped} we present the electronic band structure of \CaK\ as obtained by DFT. The result is in agreement with previous calculations of D.~Mou \textit{et. al} where the LDA  exchange-correlation functional  has been used \cite{Mou.PRL.117.277001(2016)}. Note that the degeneracy of bands along the high-symmetry paths $X\rightarrow M$ and $R\rightarrow A$, which is characteristic for other FeSC\cite{Hirschfeld_review}, is not observable in the present materials.  This feature is obviously related to the presence of the \textit{Ae}- and \textit{A}-layers surrounding the FeAs-layers and the consequent off-symmetry positions of the atoms. In addition, the K atom  in \CaK \,  lowers the underlying symmetry of the lattice, what leads to the doubling of the number of electronic bands present in the electronic structure. 

From our calculations we find the doping level of the Fe $3d$-shell $n_\text{CaK1144}=5.77$ to lie between the parant stoichiometric 122 materials, since $n_\text{Ca122}=6.07$ and $n_\text{K122}=5.43$. To compare the three systems  we have added the letter to Fig. \ref{Fig:Bands_all_Doped} and have performed a rigid band shift for \K\ and \Ca\, 
to acquire the same doping level as \CaK. In contrast to \CaK\ ,  \K\ shows
a strong 3d-like behavior of the electronic dispersion along the $\Gamma\rightarrow Z$ direction, a strongly dispersive hole pocket is observed. For \Ca\ the dispersion is more 2d-like than in \K, although the shape of the electron pockets near the $M$-point differs from that at the $A$-point, which indicates a significant corrugation of the corresponding cylinder as a function of $k_z$ in \Ca\ 
as compared to \CaK.  
Thus it seems that \CaK\ can be considered to be more 2d-like material with stronger tendency towards superconductivity as compared to the 122 counterparts.

\begin{figure}[h]
	\centering
	\includegraphics[width=7.6cm]{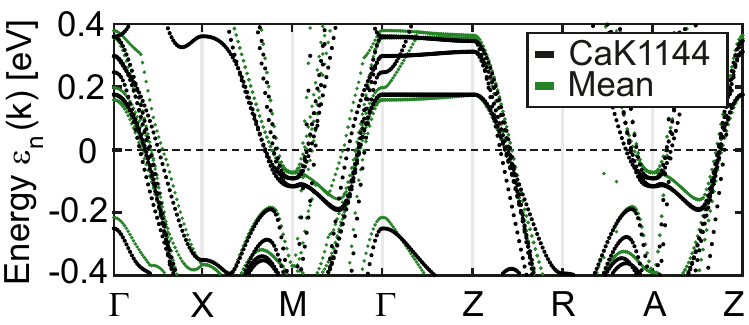}
	\caption{(color online) Electronic dispersion of 
	\CaK\ (black) compared to the mean of the band energies from \Ca\ and \K\ (green).}
	\label{Fig:bands_Middle}
\end{figure}

\begin{figure}[h]
	\centering
	\includegraphics[width=7.6cm]{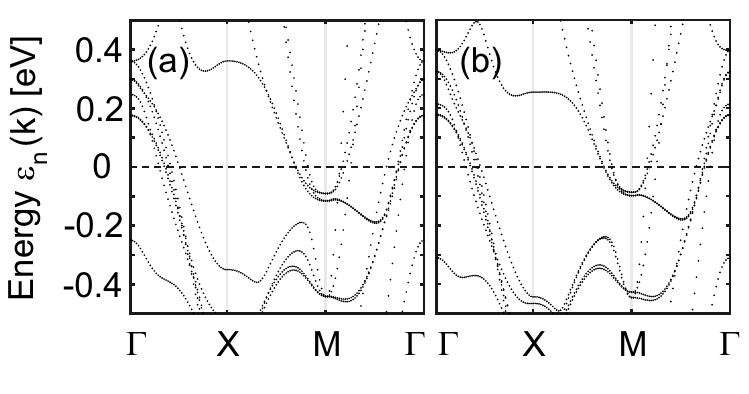}
	\caption{Influence of the off-symmetry position of the FeAs layers on the electronic band structure of \CaK. (a) shows the low-energy pat of the electronic dispersion for \CaK \, using experimental values for the atomic positions, while (b) refers to the electronic dispersion obtained for the high-symmetry Fe positions at $(0,\frac{1}{2},\pm\frac{1}{4})$ and $(\frac{1}{2},0,\pm\frac{1}{4})$. In the latter case, the distance of the As atoms is equivalent in $\pm z$ direction, using $d_\text{As-Fe}=0.105\,\mathring{\text{A}}$ as the mean of the experimental values.}
	\label{Fig:highSym}
\end{figure}

To make more visible that the structural  superposition of \Ca\ and \K\ to \CaK\ is also reflected by the band structure, we present in Fig.~\ref{Fig:bands_Middle} the comparison of \CaK\ 
electronic bands dispersions  with the ones, obtained after averaging the electronic dispersions of \Ca\ and \K\ 
systems. We focus the considerations on the low-energy properties, since they are in particular important for superconductivity.
Both structures match rather well except for the features related to the off-symmetry position of FeAs-layers in \CaK, namely the additional splitting of the energy 
bands at the $\Gamma$ point of the BZ. To see this observe
that compared to the 122 materials the FeAs-layer of \CaK\ is nearly 7.3\% out of the high-symmetry position at e.g. $(0,\frac{1}{2},\pm\frac{1}{4})$ \cite{Iyo.JACS.138.3410(2016)}. For the band structure calculations displayed in Fig. \ref{Fig:highSym} the Fe  and As atoms are forced to be at the same high-symmetry positions 
as it is the case in 122 materials. Moreover, the two different As-atoms are at equivalent positions with respect to the Fe layer. The system is therefore again described by a single Fe-As-distance, chosen as the average of the experimental values.
The resulting band-structure of the high-symmetry artificial structure turns out to be quite similar to the real one except for the slightly different position of the top of the hole bands near the $\Gamma$-point and larger splitting of the electron bands near the M-point of the BZ.

Finally let us notice that similar to the \Ca\ the experimental As positions in \CaK\ shifts by 10$\%$ in the process of the relaxation in the non-magnetic ground state of the DFT calculations. Similar to the recent finding \cite{Meier2017} we find the crystal structure is stabilized for the combined state consisting of the $(\pi,0)$ and $(0,\pi)$ ordering wave vectors in the so-called spin-vortex state. However, we are not discussing this further as we are mostly interested in constructing the effective low-energy Hamiltonian to analyse potential superconducting instabilities in \CaK\ and further employ the experimental positions for the As.   

\section{Tight-binding representation\label{sec2}}

\begin{widetext}
	
	\begin{figure}[t]
		\centering
		\includegraphics[width=16cm]{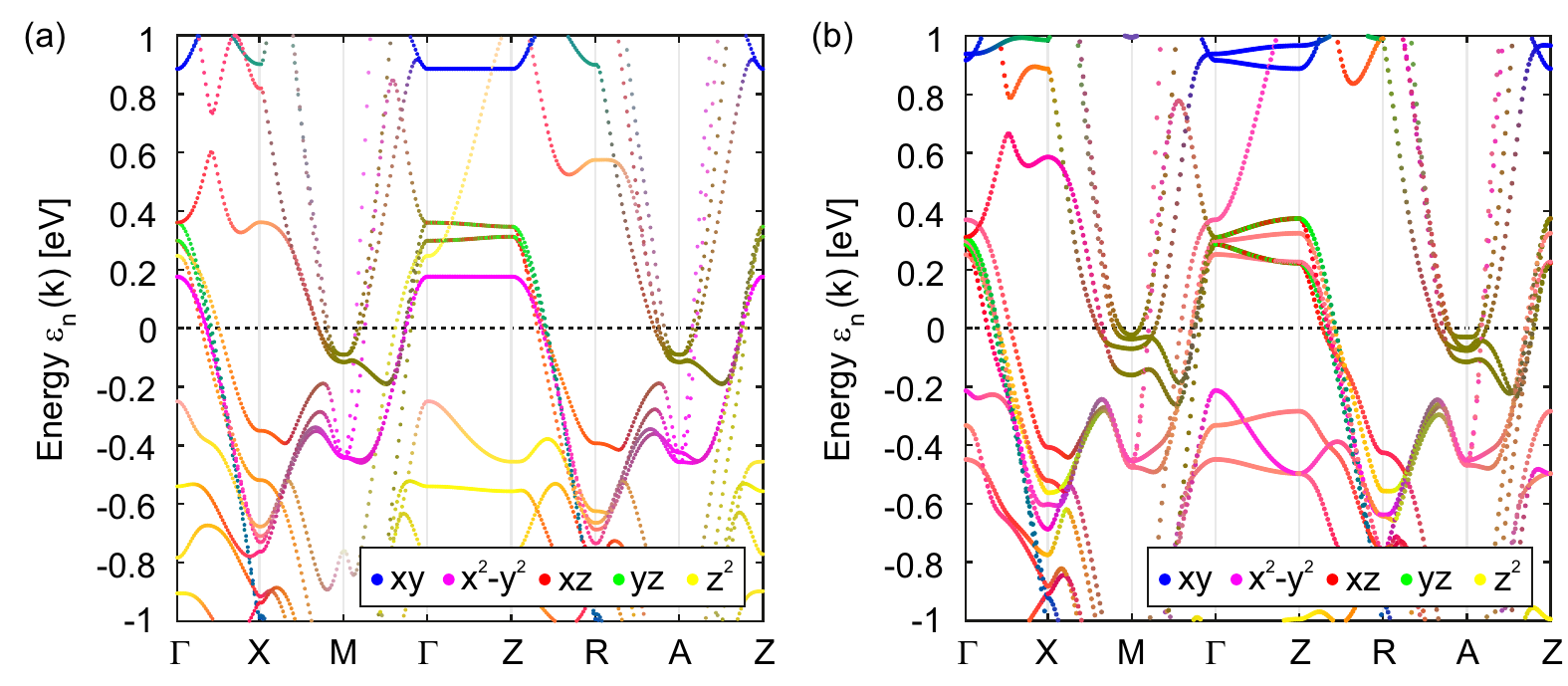}
		\caption{ (color online) Comparison between DFT results (a) and the tight-binding (b) parametrization of the energy dispersion and resulting fermiology in \CaK. Here, the $wannier90$ evaluation has been restricted to the Fe $3d$-orbital states.}
		\label{Fig:DFT_tb}
	\end{figure}
	
\end{widetext}

For a better understanding of the electronic correlation in the system, we now construct a tight-binding (tb) Hamiltonian for \CaK\ 
following the procedure made for other iron-based superconductors \cite{Eschrig.PRB.80.104503(2009),Graser.NJP.11.025016(2009),Ikeda.PRB.81.054502(2010)}. For this purpose,  we map the orbital dependent band-structure from the DFT calculations on Fe $3d$-orbital Wannier functions using the {\it wannier90} package.  
In Fig.~\ref{Fig:DFT_tb} we present the comparison between the orbitally-resolved DFT calculations (a)  and the tight-binding projection for the band structure (b) in \CaK. The corresponding terms of the tight-binding Hamiltonian are given in the Appendix.  Observe that  the lobes of the $3d$-orbitals are not elongated along the Fe-sites ($(k_x,k_y)$-coordinates), such that the orbital content is rotated by 45$^\circ$ in the $xy$-plane, forming $(k_1,k_2)$-plane coordinates. This, however, does not affect the projection on the Wannier functions. The orbital content observed within DFT is in agreement with the one calculated also previously \cite{Mou.PRL.117.277001(2016)}. One of the peculiarities of the \CaK\ electronic structure  is a strong admixture of the $d_{z^2}$-orbital to the states near the Fermi level, which is somewhat different in the \Ca\ and \K\ systems. 


The tight-binding parametrization using $3d$-Wannier orbitals of the Fe-sites, shown in Fig. \ref{Fig:DFT_tb}(b), reproduces quite well the electronic structure including the sizes of the electron and 
hole pockets, their group velocities and the orbital content. Nevertheless, one can see that for energies outside of a $0.3$\,eV 
interval near $E_F$ there are noticeable deviations between DFT and tight-binding band structures, which are related to the influences of the As $p$-orbitals in this region. 

The tight-binding Hamiltonian for \CaK\ is a good instrument to visualize and estimate the symmetries for this system, which was previously also done for the other Fe-based superconductors
\cite{Eschrig.PRB.80.104503(2009),Graser.NJP.11.025016(2009),Ikeda.PRB.81.054502(2010)}. Since we limit the $tb$ representation to the  $3d$-orbitals, we have only taken the Fe atoms into account. This leads to a block-diagonal form of the Hamiltonian caused by the different positions of the As-atoms in typical 122 FeSC material, where one As lies above (+) and another one below (-) the Fe-plane. Thus, the Hamiltonian for each FeAs-layer acquires the form
\begin{equation}
H_\text{FeSC}=
\begin{pmatrix}
H^{++} & H^{+-} \\
H^{-+} & H^{--}
\end{pmatrix}, \label{Eq:Ham_FeSC}
\end{equation}
where the superindices mark the positions of the As-atoms. Each block contains  five Fe $3d$-orbitals, which leads to the 10  orbital-resolved Hamiltonian for typical FeSC. In addition, there are symmetries for the terms in the $tb$ Hamiltonian reflecting the original crystal group symmetry of the lattice in the FeSC
\begin{equation}
\begin{aligned}
H^{++}&={H^{++}}^\dagger\\
H^{+-}&={H^{+-}}^t\\
H^{++}&={H^{--}}^*\\
H^{+-}&={H^{-+}}^*.
\end{aligned} \label{Eq:Sym_FeSc}
\end{equation}
This reduces the number of necessary $tb$ parameters \cite{Eschrig.PRB.80.104503(2009)}. With these equations the Hamiltonian gets the form
\begin{equation}
H_\text{FeSC}=
\begin{pmatrix}
H^{++} & H^{+-} \\
{H^{+-}}^* & {H^{++}}^*
\end{pmatrix}.
\end{equation}

For \CaK\ it is important to take two Fe$_2$As$_2$-layers into account due to specific symmetry of the system (see Fig. \ref{Fig:Sketch_atomPos_1144_122} (b), (d)). 
As a result the Hamiltonian requires a representation with 16 different $5\times 5$ blocks  
\begin{equation}
H_\text{1144}=
\begin{pmatrix}
H^\text{A1A1} & H^\text{A1A2} & H^\text{A1B1} & H^\text{A1B2} \\
H^\text{A2A1} & H^\text{A2A2} & H^\text{A2B1} & H^\text{A2B2} \\
H^\text{B1A1} & H^\text{B1A2} & H^\text{B1B1} & H^\text{B1B2} \\
H^\text{B2A1} & H^\text{B2A2} & H^\text{B2B1} & H^\text{B2B2}
\end{pmatrix}\text{ ,}
\end{equation}
where A/B refers to the two different Fe$_2$As$_2$-layers and the index 1/2 counts the position of the Fe-atoms (see Fig. \ref{Fig:Sketch_atomPos_1144_122}). Here, the total 
Hamiltonian for \CaK\ can be separated into four blocks each referring to a given Fe$_2$As$_2$ layer,
\begin{equation}
H_{1144}=
\begin{pmatrix}
H^\text{AA} & H^\text{AB} \\
H^\text{BA} & H^\text{BB}
\end{pmatrix} = 
\begin{pmatrix}
H^\text{AA} & H^\text{AB} \\
{H^\text{AB}}^* & {H^\text{AA}}^*
\end{pmatrix}. 
\end{equation}
In the second representation we have utilize the symmetries, shown by Eq. (\ref{Eq:Sym_FeSc}), where the $+/-$ index for the above/below As-atom has been translated to the A/B index of the FeAs-layer within the 1144 family of materials. 

Next we look at each particular block within $H^\text{AA/AB}$ terms in the Hamiltonian and analyze its symmetry transformations to reduce the number of the independent $tb$ parameters.
\begin{figure}[t]
	\centering
	\includegraphics[width=7.6cm]{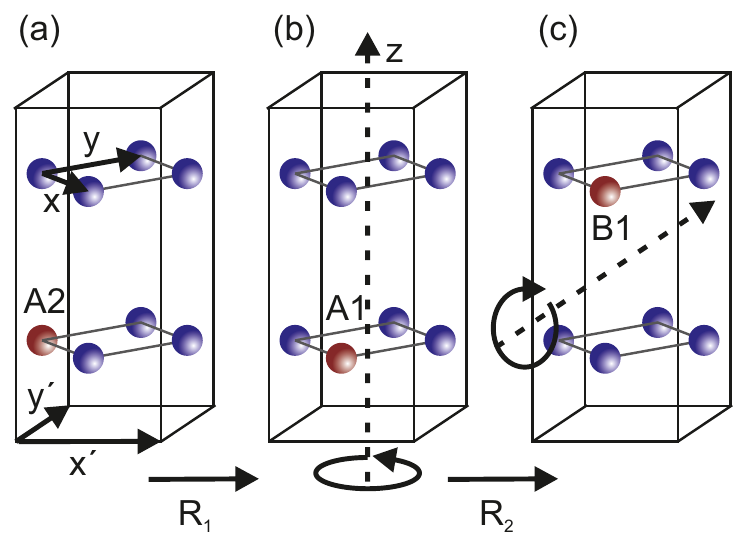}
	\caption{(color online) Introduced symmetry properties of the $tb$ Hamiltonian. The operation $R_1$ rotates the unit cell  by 90$^\circ$ along $k_z$ and $R_2$ rotates by 180$^\circ$ around the axis, shown on (c). From (a) to (b) the Fe-atom A2 is mapped on A1, from (b) to (c) A2 is mapped on B1, thus both Fe$_2$As$_2$-layers can be related by this symmetry operation. The arrows x and y refer to the basis vectors in the single Fe-unit cell ($k_x,k_y$ in the reciprocal space), while x$^{\prime}$ and y$^{\prime}$ denote the basis vectors in the 2Fe unit cell ($k_1,k_2$ in the reciprocal space).}
	\label{Fig:Sketch_rotation}
\end{figure}
For example, if the system is rotated by 180$^\circ$ around $k_2$ ($R_2$) following the initial rotation by 90$^\circ$ around the $k_z $ axis ($R_1$) the transformed coordinates read
\begin{equation}
\begin{pmatrix}
k_1 \\
k_2 \\
k_z
\end{pmatrix}
\underbrace{
	\begin{pmatrix}
	0 & 1 & 0 \\
	-1 & 0 & 0 \\
	0 & 0 & 1
	\end{pmatrix}}_{R_1}
\underbrace{
	\begin{pmatrix}
	-1 & 0 & 0 \\
	0 & 1 & 0 \\
	0 & 0 & -1
	\end{pmatrix}}_{R_2}
=
\begin{pmatrix}
k_2 \\
k_1 \\
-k_z
\end{pmatrix}.
\end{equation}
This yields a transformation matrix $U$ in the orbital basis $(d_{xy},d_{x^2-y^2},d_{xz},d_{yz},d_{z^2})$
\begin{equation}
\underbrace{
\begin{pmatrix}
1 & 0 & 0 & 0 & 0 \\
0 & -1 & 0 & 0 & 0 \\
0 & 0 & 0 & -1 & 0 \\
0 & 0 & -1 & 0 & 0 \\
0 & 0 & 0 & 0 & 1 
\end{pmatrix}}_S
\begin{pmatrix}
xy \\
x^2-y^2 \\
xz \\
yz \\
z^2
\end{pmatrix} = 
\begin{pmatrix}
xy \\
-(x^2-y^2) \\
-yz \\
-xz \\
z^2
\end{pmatrix},
\end{equation}
characterized by changes in sign and order. 
In Fig. \ref{Fig:Sketch_rotation} we illustrate the symmetry transformation $H^\text{A1A1}$ onto $H^\text{A2A2}$, which also can be written as
\begin{equation}
H^\text{A1A1}(k_1,k_2,k_z)=S^{-1}{H^\text{A2A2}(k_2,k_1,-k_z)}^*S
\end{equation}
and $H^\text{B1B1}$ can be replaced by ${H^\text{A1A1}}^*$ according to Eq.~(\ref{Eq:Sym_FeSc}). Similarly we find
\begin{align}
H^\text{A1A2}(k_1,k_2,k_z)&= S^{-1}H^\text{B2B1}(k_2,k_1,-k_z)S \nonumber \\
&= S^{-1}{H^\text{A2A1}(k_2,k_1,-k_z)}^*S. \label{eq:ReduceA} 
\end{align}
In addition, there are the symmetry relations
\begin{align}
\begin{aligned}
H^\text{A1A1}&={H^\text{A1A1}}^*\\
H^\text{A2A1}&={H^\text{A1A2}}^\dagger, 
\end{aligned}
\end{align}
which further simplify Eq. (\ref{eq:ReduceA}) to get
\begin{align}
H^\text{A1A2}(k_1,k_2,k_z)=S^{-1}{H^\text{A1A2}(k_2,k_1,-k_z)}^tS. 
\end{align}
Overall we find for the $H^\text{AB}$ blocks
\begin{align}
\begin{aligned}
H^\text{A2B2}(k_1,k_2,k_z)&= S^{-1}{H^\text{A1B1}(k_2,k_1,-k_z)}^*S\\
H^\text{A1B2}(k_1,k_2,k_z)&= S^{-1}{H^\text{A1B2}(k_2,k_1,-k_z)}^\dagger S, 
\end{aligned}
\end{align}
where in the second line we employed
\begin{align}
\begin{aligned}
H^\text{A1B1}&={H^\text{A1B1}}^t\\
H^\text{A2B1}&={H^\text{A1B2}}^t, 
\end{aligned}
\end{align}
which stems from the symmetry $H^\text{AB}={H^\text{AB}}^t$. Moreover, the chosen symmetry operations (rotations $R_1$ and $R_2$) are necessary to present the tight-binding Hamiltonian in the proper hermitian form as only the upper triangle of the blocks are sufficient to construct the full Hamiltonian. As a result only four blocks $H^\text{A1A1}$, $H^\text{A1A2}$, $H^\text{A1B1}$ and $H^\text{A1B2}$ appear now to have an independent form.

Observe that the 122 and 1144 tight-binding Hamiltonians are similar except that 
the higher symmetry of the 122 systems yields identical hopping matrix elements along the $k_1$ and 
$k_2$ directions,  see Eq.~(\ref{Eq:Ca122A1A1}) in the Appendix, which is not the case for 
\CaK. Moreover, the hopping for the $d_{xz}$ and $d_{yz}$-orbitals of 122 are interchanged, i.e.  $(k_x, k_y) \rightarrow (k_y,k_x)$, except for the sign. The special symmetries of the 122 materials are  also present in the hoppings  of A1B1 and A1B2. Here, the hoppings  along $k_z$ are identical to the hoppings in the ($k_1$,$k_2$)-plane. This is the result of the gliding symmetry present in 122 and again absent in \CaK. Interestingly, this is one of the reasons for the stronger quasi-two-dimensionality in \CaK\ as compared to the 122 structures. In its turn the quasi-two dimensional electronic structure allows a relatively straightforward analysis of the superconducting instabilities, which we do in the next section.

\section{Hubbard-Hund Hamiltonian and Superconducting Instabilities \label{sec3}}

Based on the single-particle low-energy Hamiltonian, we analyze the superconducting instabilities in \CaK\ by employing the random phase approximation (RPA) within the leading angular harmonics approximation (LAHA) for the Hubbard-Hund Hamiltonian \cite{Maiti2011a,Maiti.PRB.85.014511(2012),Ahn.PRB.89.144513(2014)}. This Hamiltonian is given by
\begin{eqnarray}
\lefteqn{H_{\text{int}} =  \frac{U}{2}\sum_{\substack{i,s\\\sig}}n_{i s\sig}n_{i s\sigb}+ \frac{U^\prime}{2}\sum_{\substack{i,s\neq t  \\\sig,\sigs}}n_{i s\sig}n_{i t \sigs}}&& \nonumber \\
&& - J\sum_{\substack{i, s \neq t}} S_{i s}\cdot S_{i t}+\frac{J^\prime}{2}\sum_{\substack{i, s \neq t\\\sig}}d_{i s \sig}^\dag d_{i s \sigb}^\dag d_{i t \sigb}d_{i t \sig}\;.
\end{eqnarray}
where $s$ and $t$ here refer to the orbital indices. 
The other symbols label the intra-orbital Hubbard interaction $U$, the inter-orbital Hubbard interaction $U'$, the inter-orbital
exchange $J$, and the pair hopping term $J'$. We assume that the interactions originate from a single two-body term with spin rotational invariance, i.e.
$J' = J$ and $U' = U - 5 J/2$.  This leaves $U$ and $J$ as the only two parameters in the problem.  Here, $U$ defines the overall magnitude of the pairing interaction, while the structure of the superconducting gap depends on the single parameter $J/U$, which we will vary. 

In order to study the BCS-type superconductivity in the band representation, the interactions need to be rewritten in terms of the band operators. The transformed interactions describe the repulsion between band fermions and acquire a momentum (angular) dependence because of the underlying orbital structure. We use the LAHA formalism  to solve the BCS-type gap equation ~\cite{Maiti2011b,Maiti.PRB.85.014511(2012),Ahn.PRB.89.144513(2014)} by separating each interaction between fermions on $i$ and $j$ pockets
$(i,j=h_1,\dots,h_6\text{ and }e_1,\dots,e_4)$ 
into $s-$wave, $d_{x^2-y^2}$, and $d_{xy}$ channels. As the superconducting gap symmetry in \CaK\ belongs to the $A_{1g}$-symmetry representation \cite{Mou.PRL.117.277001(2016)} we restrict ourselves only to these solutions to make the gap equation tractable semi-analytically and to be able to follow the gap evolution upon changing the parameters.  

We apply LAHA, solve for superconductivity and vary the parameters of the underlying model to see whether the solutions that we find are stable with respect to the variation of the interactions.
Choosing this approach, orbital effects and spin-fluctuations (SF) will determine the strength of the effective interactions that describe the scattering of a Cooper pair between different Fermi surface pockets and will determine the gap structure.  We  extend the computational procedure described in \cite{Maiti2011a} to the fermiology of Ca1144 to compare possible superconducting s-wave spin-singlet states.

The transformation for the repulsive on-site Coulomb interactions is a consequence of the diagonalization of the kinetic part of the Hamiltonian, which is given in tight-binding representation (Sec.~\ref{sec2}).  In the notation of \cite{Maiti2011a,Ahn.PRB.89.144513(2014)}, the effective interactions are named $\Gamma_{ij}(\mathbf{k},\mathbf{k^\prime})$.

Once the problem is reduced to Cooper-pairs near the Fermi level, i.e. once the momenta are constrained to the contours of the Fermi surface pockets in the \mbox{$k_x\text{-}k_y$-plane}, the momentum dependence can be decomposed into leading angular harmonics. The Fermi surface pockets are quasi 2D cylinders (Fig.~\ref{fig:FScuts}), which are periodic in \mbox{$k_z$-direction}. Therefore, it is also possible to find the leading harmonics of the $k_z$ momentum dependence of the effective interactions and the gap function. However, we note that the first harmonic, i.e., a constant, is already a good approximation. In other words, we solve the 3D BCS-type gap equations, but only give the 2D result as first order approximation, since the gap function is only weakly dispersing in \mbox{$k_z$ direction}. 

Since the Fermi surface topology of \CaK\ consists of electron and hole pockets of a relatively small radius and shows a weak $k_z$ dispersion, we can apply LAHA,  where the Cooper-pair scattering can be distinguished between intraband (small {\bf q}) and interband (large {\bf q}) ones. Interestingly, in contrast to the situation of some ferropnictides like LiFeAs, where a strong orbital differentiation for some of the pockets takes place, here the orbital character of $d_{xz}$, $d_{x^2-y^2}$, and $d_{yz}$ is nearly equally distributed between electron and hole pockets, which is also seen in Fig. 5.
Furthermore, the hole bands also acquire some $d_{z^2}$ character, which is often  also the case in other iron-based superconductors.
Its influence was usually ignored in effective models, which consider a $d_{xz}$, $d_{x^2-y^2}$, and $d_{yz}$ orbital character only \cite{Ahn.PRB.89.144513(2014)}. 


\begin{widetext}
	
	\begin{figure}[H]
		\centering
		\begin{minipage}{0.49\textwidth}	
			\centering			
			{\includegraphics[height=0.8\textwidth,trim=2mm 2mm 10.25mm 18mm,clip]{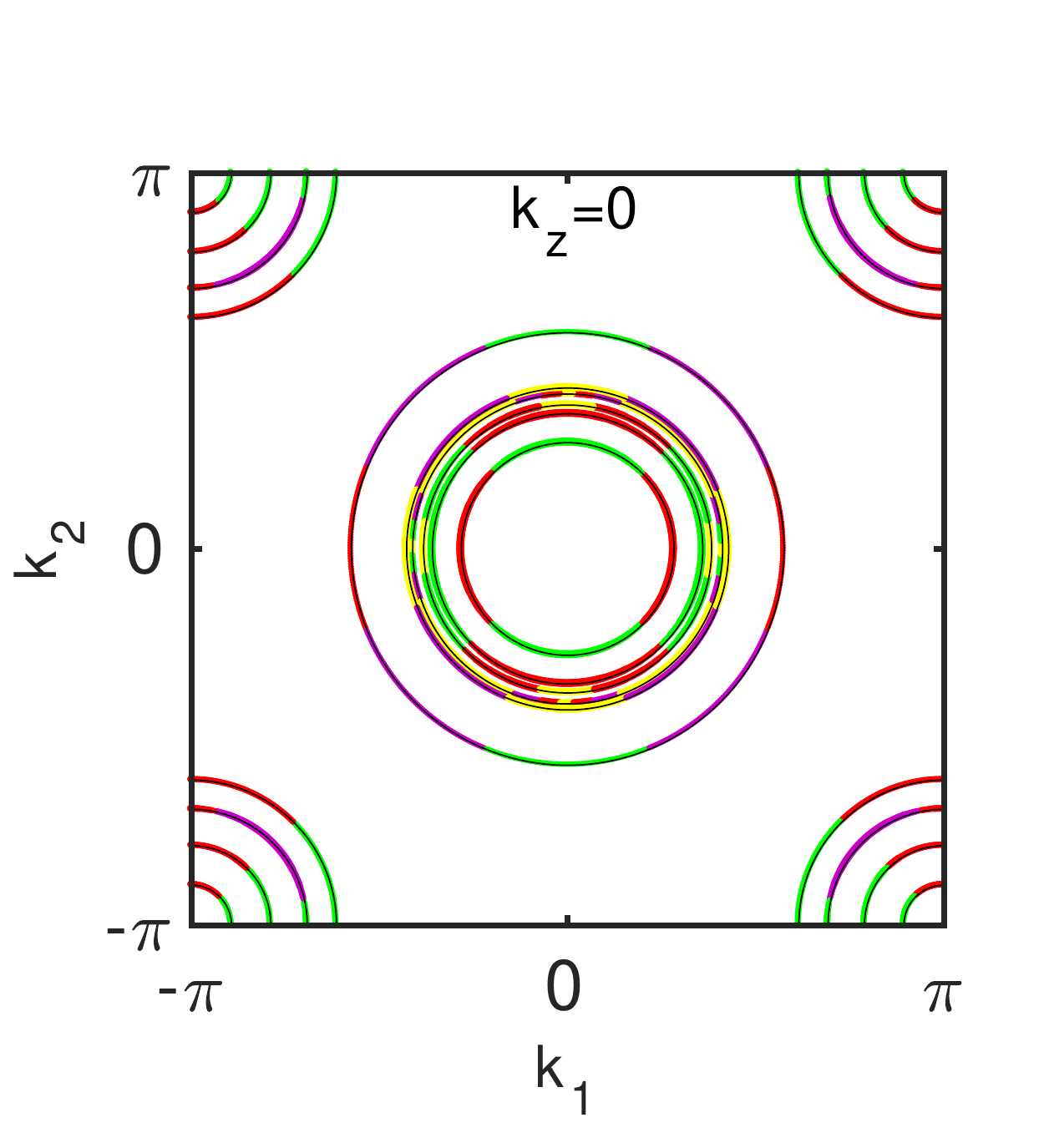}} %
			
		\end{minipage}
		\begin{minipage}{0.49\textwidth}
			\centering
			{\includegraphics[height=0.84\textwidth,trim=12.5mm 4.5mm 0mm 18mm,clip]{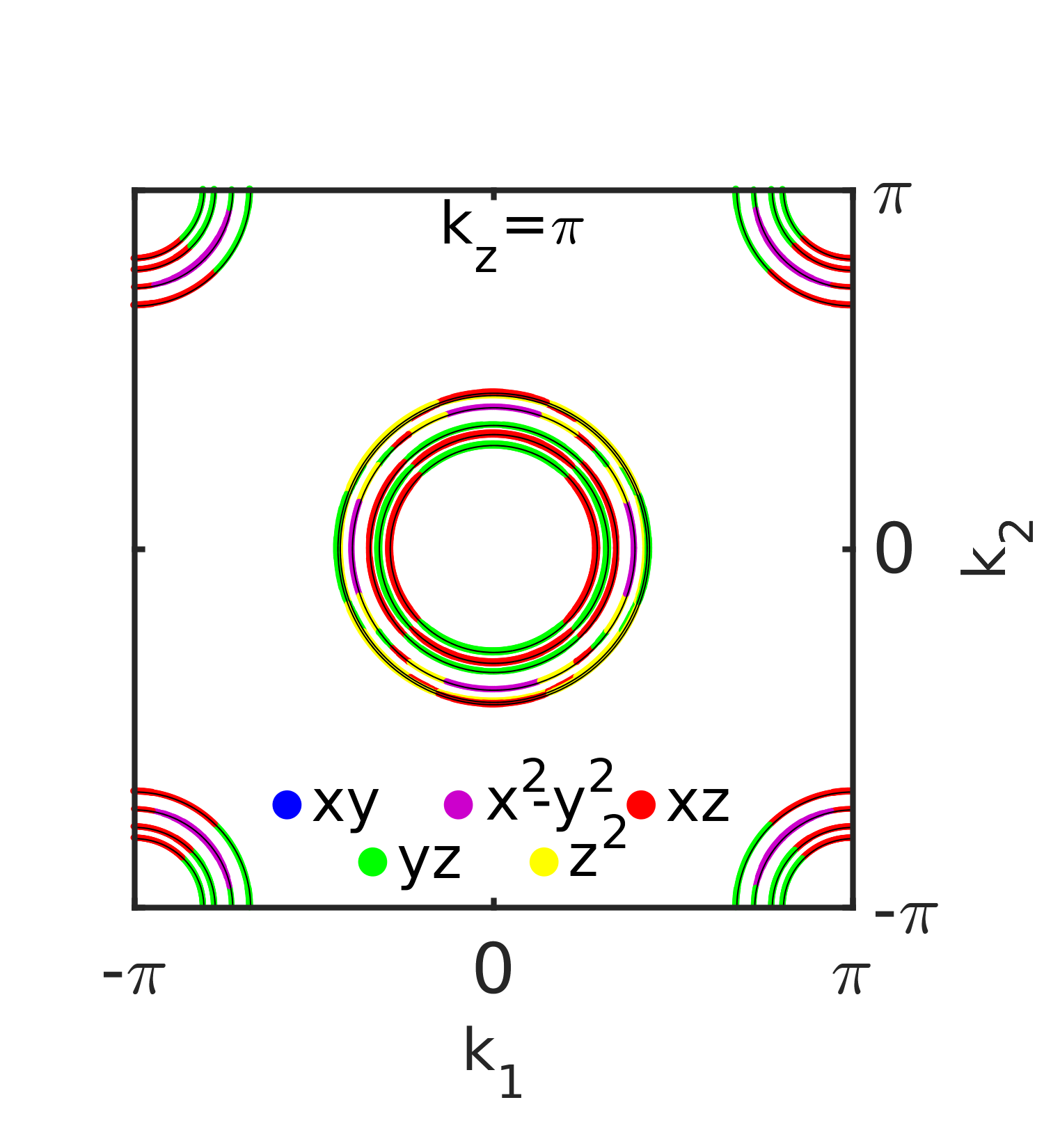}} %
		\end{minipage}		
		\caption{Fermi surface structure of the \CaK\ for two different $k_z$ cuts with 6 hole and 4 electron Fermi surface pockets. The electronic states at the Fermi level are colored regarding their largest contribution from the iron \mbox{$3d$-orbitals}.}\label{fig:FScuts}
		\label{fig:FScuts}
	\end{figure}

\end{widetext}
In the next step we compute the spin response functions within  RPA, following the original proposal in Ref.~\cite{Graser.NJP.11.025016(2009)}.
The spectral representation of the Green's function is given as
\begin{equation}
G_{sp} (k,i\omega_n) = \sum_\mu \frac{a_\mu^s(k) a_\mu^{p*}(k)}{i \omega_n - E_\mu(k)}.
\end{equation}
Here, the matrix elements $a_\mu^s(k) = \langle s | \mu k \rangle$
connect the orbital indices $s,p$ and the band index $\mu$  and are the
components of the eigenvectors resulting from the diagonalization
of the tight-binding Hamiltonian. We
find the non-interacting susceptibilities
\begin{eqnarray}
\chi_{st}^{pq} (q,\omega) & = & - \frac{1}{N} \sum_{k,\mu\nu} \frac{a_\mu^s(k) a_\mu^{p*}(k) a_\nu^q(k+q) a_\nu^{t*}(k+q)}
{ \omega + E_\nu(k+q) - E_\mu(k) + i 0^+} \nonumber \\
& & \times \left[ f(E_\nu(k+q)) - f(E_\mu(k)) \right]. 
\end{eqnarray}
The RPA expression for the spin susceptibility is given in the form of the Dyson-type
equation
\begin{equation}
(\chi_{1}^{\rm RPA})_{st}^{pq} = \chi_{st}^{pq} +
(\chi_{1}^{\rm RPA})_{uv}^{pq} (U^s)_{wz}^{uv} \chi_{st}^{wz},
\end{equation}
where the summation is assumed over repeated indices. Here the non-zero
components of the matrices of the spin dependent interaction $U^s$ are given as
\begin{align*}
(U^s)_{aa}^{aa} &= U,& (U^s)_{bb}^{aa} &= \frac{1}{2}J,\\
(U^s)_{ab}^{ab} &= \frac{1}{4} J + U^{\prime},& (U^s)_{ab}^{ba} &= J',
\end{align*}
where $a\neq b$.

\begin{widetext}
	
\begin{figure}[H]
	\centering
	\begin{minipage}{0.49\textwidth}	
		\centering			
		{\includegraphics[height=0.84\textwidth,trim=0mm 0mm 8mm 8mm,clip]{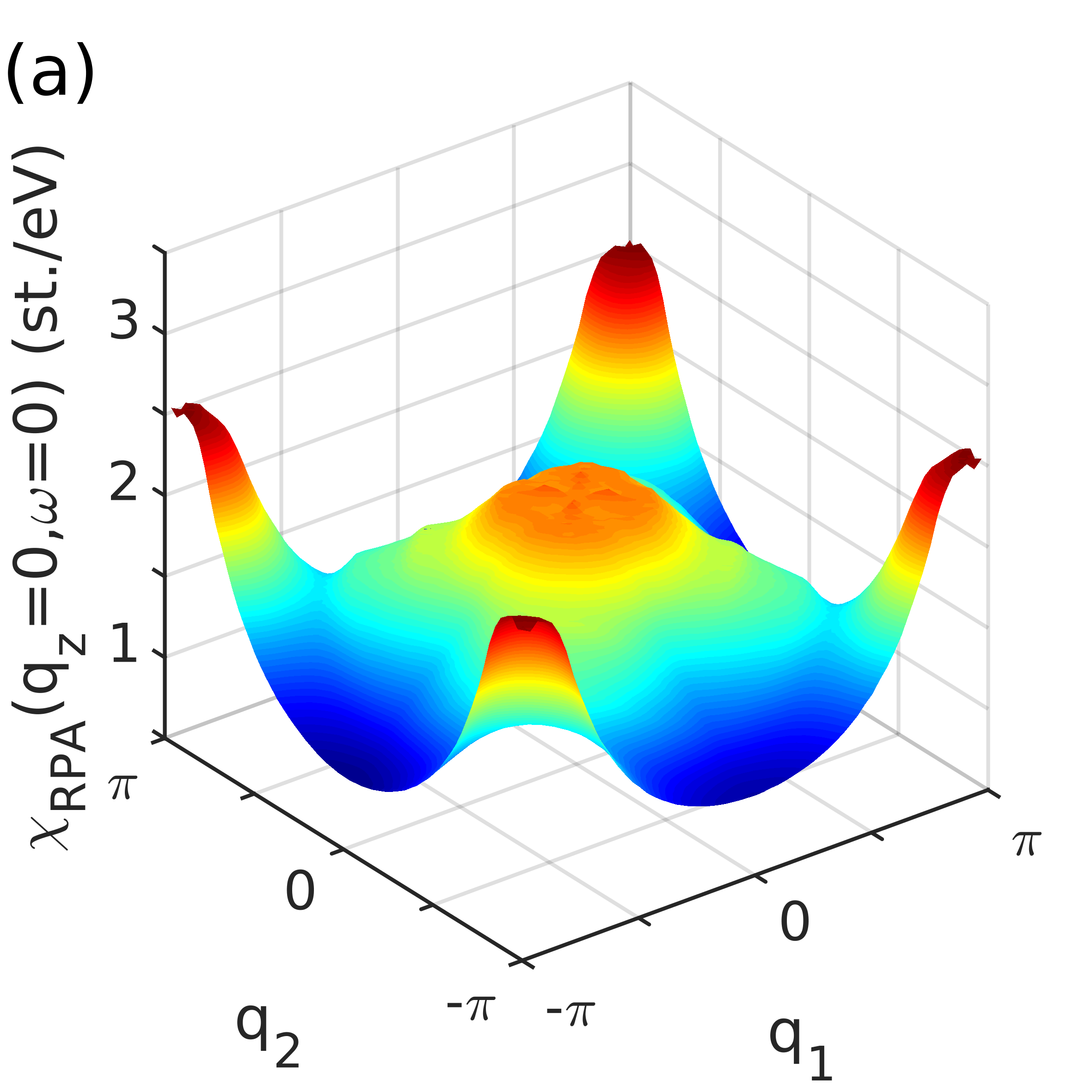}} %
	\end{minipage}
	\begin{minipage}{0.49\textwidth}
		\centering
		{\includegraphics[height=0.84\textwidth,trim=0mm 0mm 8mm 8mm,clip]{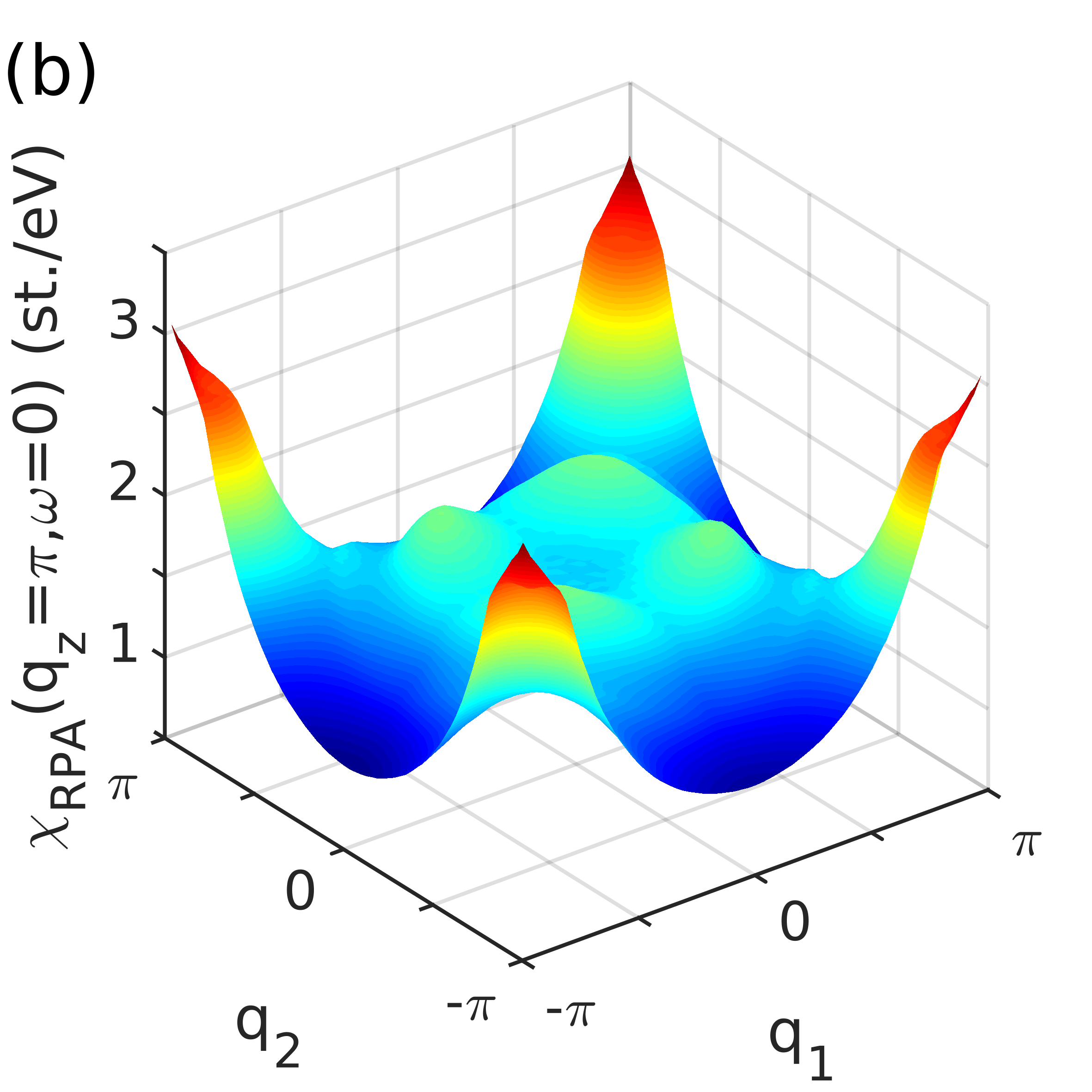}} %
	\end{minipage}		
	\caption{Multi-orbital spin susceptibility within RPA approximation at zero temperature for $q_z=0$ (a), $q_z=\pi$ (b), Coulomb repulsion $U=1.5\text{eV}$ and Hund $J=0.1U$.}\label{fig:chiRPA}	
\end{figure}

\end{widetext}

The results of the calculations are shown in Fig. \ref{fig:chiRPA} for two different $k_z$ cuts. One could clearly see that even for the moderate values of the interaction the susceptibilities are enhanced for the antiferromagnetic wave vector
reflecting the nesting of the hole and electron dispersions, $E_e({\bf k}) = -E_h({\bf k+Q})$ for ${\bf Q}=(\pi,\pi)$ in the
two Fe BZs.
Furthermore, one clearly sees the peaks at smaller wave vectors, resulting from the scattering between the hole bands.

The spin fluctuation component of the interaction is obtained by summing up
second and higher-order ladder diagrams in the orbital
formalism. The total interaction is then converted from orbital to
band basis by dressing it by matrix elements associated with the
hybridization of Fe orbitals \cite{Maiti2011a}. Aiming at the analysis of superconductivity, the end result of this procedure
is the
effective BCS-type Hamiltonian in the band description
\beq
{\cal H} = \sum_{i,{\bf k}} \epsilon_{i} ({\bf k}) c^\dagger_{i {\bf k}} c_{i {\bf k}} +
\sum_{i,j, {\bf k}, {\bf k}'} \Gamma_{ij} ({\bf k},{\bf k}') c^\dagger_{i {\bf k}} c^\dagger_{i -{\bf k}} c_{j {\bf k}'}c_{j -{\bf k}'}.
\label{1}
\eeq
The quadratic term describes low-energy excitations near hole and
electron Fermi surface sheets, labeled by $i$ and $j$, and the interaction term describes
the scattering of a pair $(k\uparrow, -k\downarrow)$ on the pocket $i$ to
a pair $(-k'\uparrow, k'\downarrow)$ on the pocket $j$. The
effective singlet interaction $\Gamma_{ij}({\bf k},{\bf k}')$ is then given by
\begin{eqnarray}
{\Gamma}_{ij} ({\bf k},{\bf k}') & = & \sum_{s,t,p,q} a_{\nu_i}^{t*}(-{\bf k})
a_{\nu_i}^{s*}({\bf k})
\mathrm{Re}\left[{\Gamma}_{st}^{pq} ({\bf k},{\bf k}',0) \right] \nonumber \\
& \times & a_{\nu_j}^{p}({\bf k}')  a_{\nu_j}^{q}(-{\bf k}'),
\end{eqnarray}
with
\begin{eqnarray}
{\Gamma}_{st}^{pq} ({\bf k},{\bf k}',\omega) &=& \left[\frac{3}{2} U^s
\chi_1^{\rm RPA}  ({\bf k}-{\bf k}',\omega) U^s   \right]_{ps}^{tq}.
\label{eq:fullGamma}
\end{eqnarray}
Here we drop the orbital
(charge)-fluctuation contribution as described in
Ref.~\cite{Maiti2011a}.

In Tables I-III we present exemplarily for $U=1.5$\,eV 
and $J/U=0.1$ the results of the LAHA projection for the intraband and interband interaction for the 10 pockets in \CaK. As experimentally the angular variation of the superconducting gap on each pockets is found to be negligible \cite{Mou.PRL.117.277001(2016)}, we restrict ourselves to the constant superconducting gaps on each Fermi surface pocket and constant interactions. On average we find that the Cooper-pairing interactions are stronger between electron and hole bands than between the bands of the same character, i.e. hole-hole or electron-electron bands, which is a result of the spin fluctuations enhancement of the Cooper-pairing interaction. 
Nevertheless, there is still strong intraband repulsion for some of the bands like 
$e_2$ or $h_4$.

\begin{table}[H]
	\centering
	\begin{tabularx}{0.4\textwidth}{XXXXXXX} 
		\toprule
		$\Gamma_{ij}$&	$h_1$	&	$h_2$	&	$h_3$	&	$h_4$	&	$h_5$	&	$h_6$	\\
		\midrule
		$h_1$	&	0.61	&	0.61	&	0.12	&	0.08	&	0.09	&	0.08	\\
		$h_2$	&			&	0.58	&	0.11	&	0.06	&	0.08	&	0.07	\\
		$h_3$	&   		&			&	0.49	&	0.07	&	0.32	&	0.33	\\
		$h_4$	&  			&			&			&	1.10	&	0.08	&	0.09	\\
		$h_5$	&   		&	 		&			&			&	0.42	&	0.44	\\
		$h_6$	&			&			&			&			&			&	0.39	\\
		\bottomrule
	\end{tabularx}
	\caption{Effective interactions, scattering of Cooper-pairs from one hole pocket to another.}
\end{table}

\begin{table}[H]
	\centering
	\begin{tabularx}{0.4\textwidth}{XXXXX} 
		\toprule
		$\Gamma_{ij}$&	$e_1$	&	$e_2$	&	$e_3$	&	$e_4$	\\
		\midrule
		$e_1$	&	0.42	&	0.06	&	0.30	&	0.35	\\
		$e_2$	&			&	1.08	&	0.04	&	0.04	\\
		$e_3$	&   		&			&	0.75	&	0.28	\\
		$e_4$	&  			&			&			&	0.66	\\
		\bottomrule
	\end{tabularx}
	\caption{Effective interactions, scattering of Cooper-pairs from one electron pocket to another.}
\end{table}

\begin{table}[H]
	\centering
	\begin{tabularx}{0.4\textwidth}{XXXXXXX} 
		\toprule
		$\Gamma_{ij}$&	$e_1$	&	$e_2$	&	$e_3$	&	$e_4$	\\

		\midrule
		$h_1$	&	0.87	&	0.10	&	0.87	&	0.46	\\
		$h_2$	&	0.47	&	0.19	&	0.42	&	0.31	\\
		$h_3$	&   0.57	&	0.12	&	0.53	&	0.54	\\
		$h_4$	&  	0.42	&	0.46	&	0.54	&	0.33	\\
		$h_5$	&   0.50	&	0.68 	&	0.37	&	0.83	\\
		$h_6$	&	0.12	&	1.50	&	0.10	&	0.13	\\
		\bottomrule
	\end{tabularx}
	\caption{Effective interactions, scattering of Cooper-pairs from hole pockets to electron pockets.}
\end{table}

\begin{figure}[t]
	\centering
	\includegraphics[width=7.6cm]{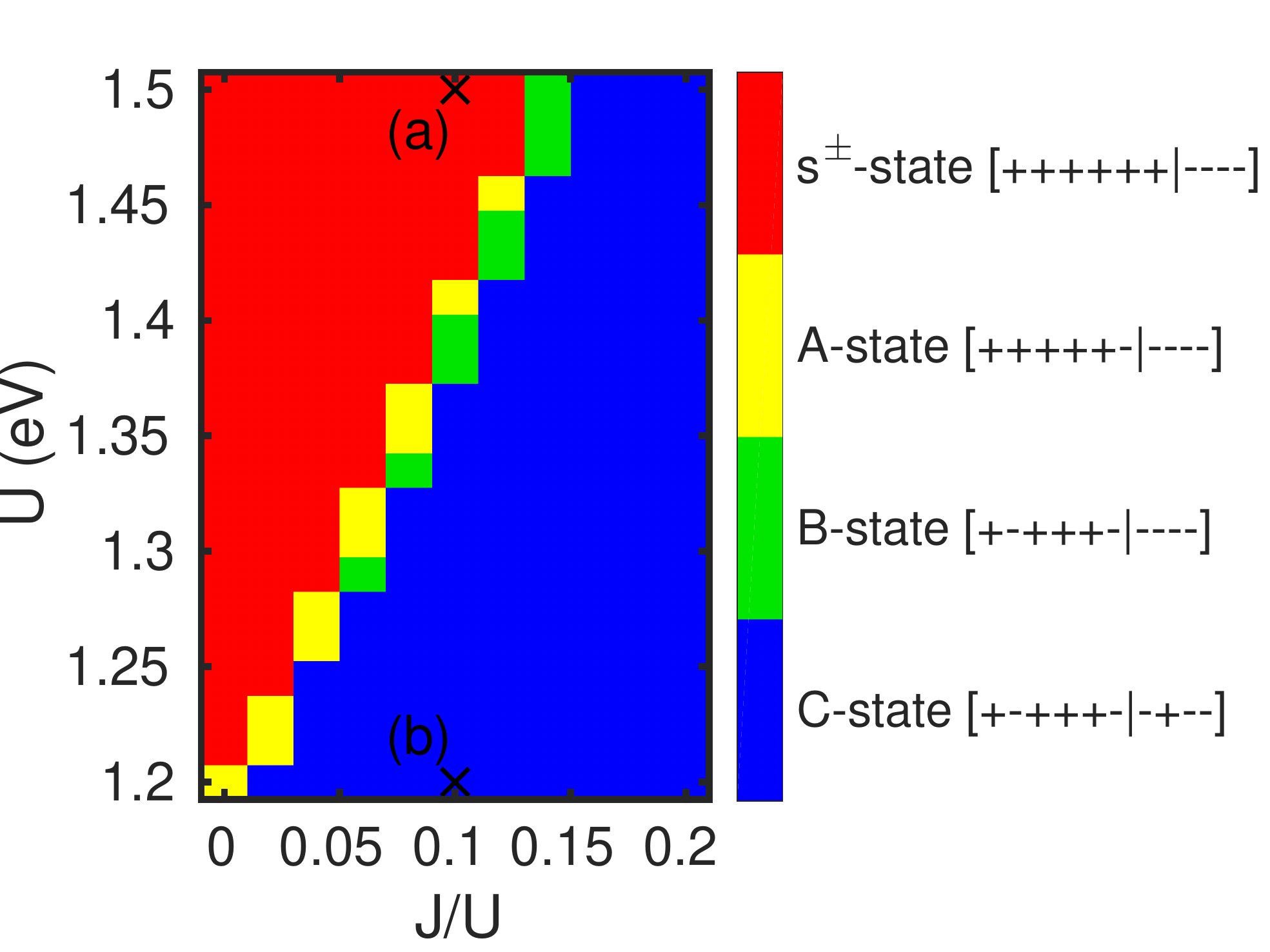}
	\caption{ Sign-structure of the leading solution in the $A_{1g}$ channel of the linearized BCS-type gap equations as a function of 
		$U$ and the $J/U$ ratio. 
		The SC gap at the six hole and four electron Fermi surface pockets can either have a positive or negative sign, which is denoted by the vector $[h_1\dots h_6,e_1\dots e_4]$. We find that spin-fluctuations at a antiferromagnetic wave vector determine the sign-structure of the superconducting gap and lead to a conventional s$^\pm$-wave state. 
		In particular, this scenario is likely for $U\gg J$. The crosses refer to the particular ratio of the gaps presented in Fig.~\ref{fig:comexp}.}
	\label{Fig:Lahadiagram}
\end{figure}

We substitute the obtained interactions into the multiband version of the coupled linearized BCS equation, which has the form 
\begin{widetext}
\begin{align}
(-\lambda)%
\left(\begin{matrix}
\Delta_{h_1}\\
\vdots\\
\Delta_{h_6}\\
\Delta_{e_1}\\
\vdots\\
\Delta_{e_4}\\
\end{matrix}\right)%
&= 
L\left(\begin{matrix}
\Gamma_{h_1h_1}        &    \hdots    & \Gamma_{h_1h_6}    &    \Gamma_{h_1e_1}    &    \hdots & \Gamma_{h_1e_4}\\
\vdots                &    \ddots    &    \vdots &    \vdots            &    \ddots    & \vdots        \\
\Gamma_{h_6h_1}        &    \hdots    & \Gamma_{h_6h_6}    &    \Gamma_{h_6e_1}    &    \hdots & \Gamma_{h_6e_4}\\
\Gamma_{e_1h_1}        &    \hdots    & \Gamma_{e_1h_6}    &    \Gamma_{e_1e_1}    &    \hdots & \Gamma_{e_1e_4}\\
\vdots                &    \ddots    &    \vdots &    \vdots            &    \ddots    & \vdots        \\
\Gamma_{e_4h_1}        &    \hdots    & \Gamma_{e_4h_6}    &    \Gamma_{e_4e_1}    &    \hdots & \Gamma_{e_4e_4}\\
\end{matrix}\right)%
\left(\begin{matrix}
\Delta_{h_1}\\
\vdots\\
\Delta_{h_6}\\
\Delta_{e_1}\\
\vdots\\
\Delta_{e_4}\\
\end{matrix}\right).%
\end{align}
\end{widetext}
We solve it numerically for various strength of the intraorbital on-site Coulomb repulsion $U$ and Hund coupling $J$, following the original procedure \cite{Ahn.PRB.89.144513(2014)}. The resulting phase diagram of the leading instabilities in the $A_{1g}$ channel is shown in Fig. \ref{Fig:Lahadiagram}. Observe also that once the spin fluctuations, which enhance the interaction between electron and hole bands, are included the $d_{x^2-y^2}$-wave ($B_{1g}$) and $d_{xy}$-wave ($B_{2g}$) symmetry solutions appear to loose against $s$-wave ones.

As a consequence, the phase diagram is dominated by two types of solutions: The first one could be regarded as a conventional $s^{\pm}$-wave in which the order parameter changes sign between hole and electron pockets. This symmetry is promoted by the strong intraorbital antiferromagnetic spin fluctuations, enhanced by the Coulomb repulsion, $U$. At the same time for larger $J/U$ ratios and smaller $U$ the sign structure of the superconducting gaps is distributed in a more sophisticated way between the pockets and involves an additional sign change within hole and electron pockets. This is mainly due to the fact that for increased $J/U$ ratio some of the interband interactions change sign and become weakly attractive. In addition the spin fluctuation enhancement is weaker for smaller $U$ values. This modifies the balance for the conventional $s^{\pm}$-wave state and promotes states where the order parameter also changes sign within electron or hole pockets. We obtain the most stable solution of this type, when at least one of the hole and one of the electron pockets changes its sign with respect to their counterparts, which we denote the $C$-state solutions.
	
\begin{widetext}
	
	\begin{figure}[H]
		\centering
		\begin{minipage}{0.45\textwidth}	
			\centering			
			{\includegraphics[height=0.65\textwidth,trim=0mm 0mm 8mm 8mm,clip]{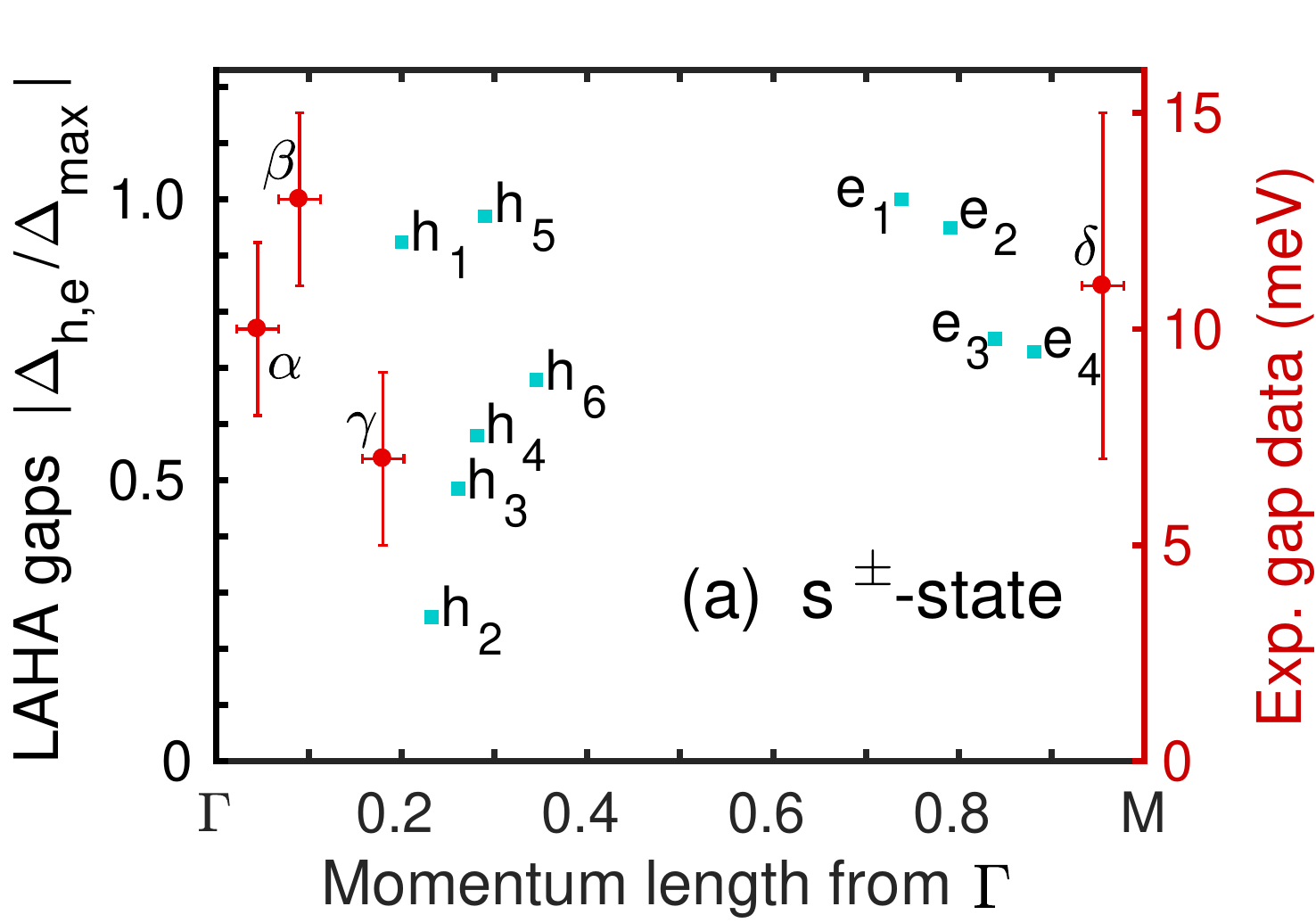}} %
		\end{minipage}
		\begin{minipage}{0.45\textwidth}
			\centering
			{\includegraphics[height=0.65\textwidth,trim=0mm 0mm 0mm 8mm,clip]{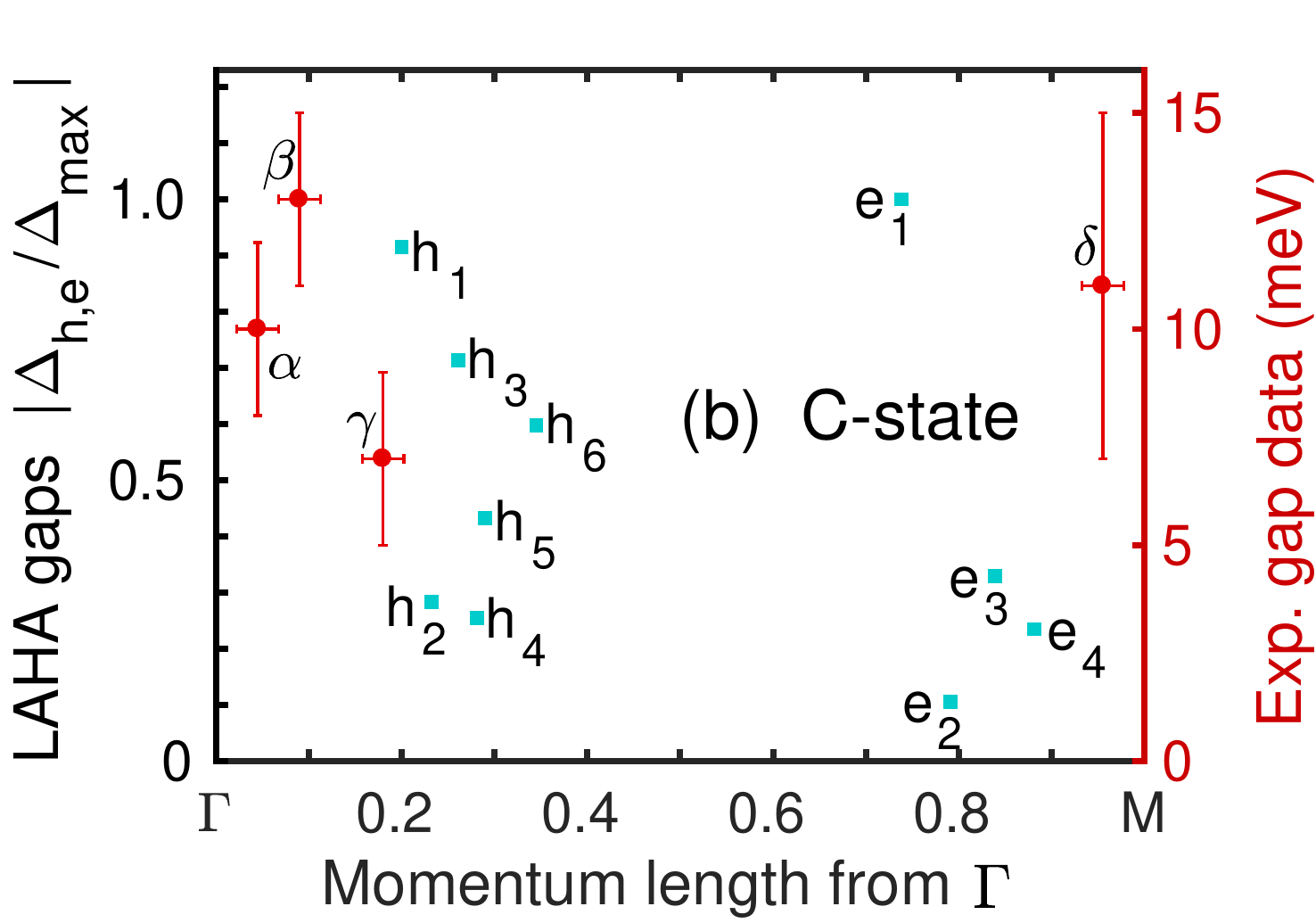}} %
		\end{minipage}		
		\caption{Mean values of the s$^\pm$-wave superconducting gap evaluated on the different Fermi surface pockets for two different values of $U=1.5$ eV (a)-the conventional $s^{\pm}$-state and $U=1.2$ eV (b)-the so-called $C$-state and $J/U=0.1$. The red points refer to  experimental values \cite{Mou.PRL.117.277001(2016)}.  }\label{fig:comexp}	
	\end{figure}
		
\end{widetext}

We note that our theory contains more bands than seen in ARPES experiments \cite{Mou.PRL.117.277001(2016)}. This could be partially due to the near degeneracy of some of the hole pockets as well as the electron ones and their similar orbital content, which prevents their straightforward identification in the ARPES experiments. In addition we also find that the electron and hole bands experience the so-called red/blue shift, i.e. their $k_F$ values are smaller than those found in DFT calculations. This is a general feature observed in many FeSCs \cite{Charnukha2015} and is believed to arise from the effect of the strong interband repulsion between electron and hole bands \cite{Ortenzi2009}. 
Nevertheless, we do not expect that this changes the results of the theoretical calculations as the density of states in two dimensions does not depend  on the radius of the Fermi surface sheet and therefore the interaction strength, determined from LAHA, remains unaffected. In addition we also checked that the orbital character of the bands does not change once the Fermi radius of each of the bands is reduced. Furthermore, the actual value of $E_F$, counted from the bottom of the electron band, or the top of the hole bands is still much larger than the superconducting gap values, which legitimises  the use of the standard multiband BCS theory.  

Experimentally, also the sizes of the superconducting gap on some of the Fermi surface sheets, namely on three hole pockets and one electron pocket, were measured by ARPES \cite{Mou.PRL.117.277001(2016)}. In particular, it was found that the largest gap appears for the electron and hole pockets that are nearly perfectly nested, which was  interpreted in favour of the conventional $s^{\pm}$-wave gap. In Fig. \ref{fig:comexp} we present the results of our calculations from the phase diagram, shown in Fig. \ref{Fig:Lahadiagram} for the $s^{\pm}$-wave (a) and the $C$-state (b) that match best the experimental values \cite{Mou.PRL.117.277001(2016)}. We observe that the sizes of the gaps for the $s^{\pm}$-wave states is closer to the values found experimentally. For the $C$-state it turns out that there is a larger distribution of the gaps on the electron Fermi surface cylinders, which would be reflected in the near-nodal behaviour of the quasiparticle excitations on some of the electron pockets, which is not seen up to now. The conventional $s^{\pm}$-state 
shows values which are quite consistent with those found experimentally. This is further supported by the proximity to the nesting of the electron and hole bands. Nevertheless, phase sensitive experiments are needed to confirm this state in \CaK.   

Unfortunately, the three dimensional character of the bands in \Ca\ or \K\ does not allow an immediate application of the LAHA approach to theses 122 systems.  However, on general grounds one would expect the stronger tendency towards $s^{\pm}$ superconductivity in \CaK\ due to the revealed two-dimensional character of the electronic bands in the latter. In principle, we expect that moving the layers in the non-symmetric position should enhance the two-dimensionality of the system, allowing easier formation of unconventional superconductivity.

\section{Conclusion\label{sec4}}

To conclude, we investigated the electronic structure in \CaK\ using density functional theory. We systematically compared the electronic structure of the 1144 and the 122 materials and analyze the influences of the off-symmetry positions of the FeAs-layer in \CaK. In particular, we find that \CaK\ could be well described as a doped 122 system with some caveat, introduced by the off-symmetry position of the $Fe_2 As_2$ layers in \CaK. Among them is a near-degeneracy of several hole bands near the Fermi level and their multiple orbital content, consisting of $d_{yz}$, $d_{xz}$, $d_{xy}$ ($d_{x^2-y^2}$), and $d_{z^2}$ orbitals. One of the most important consequences, however, is the actual two-dimensional electronic structure in \CaK\ as compared to 122 materials, which arise due to the absence of the gliding symmetry in \CaK. 

We develop the low-energy description of this system by projecting the DFT electronic structure on the tight-binding (TB) Hamiltonian based on the Fe $3d$-orbitals only and discuss the different symmetries within the system. We then use this Hamiltonian and Hubbard-Hund intrasite interaction terms as a basis to investigate potential superconducting instabilities in \CaK. The nesting between strongly two-dimensional electron and hole bands  supports strongly the $A_{1g}$ symmetry representation for the superconducting gap with most likely $s^{\pm}$-wave symmetry where the gap magnitude changes phase between electron and hole pockets. For the increased  Hund coupling, $J$  other solutions are also possible and it remains to be determined experimentally which particular phase structure the superconducting order parameter  has in \CaK.

\section*{Acknowledgments}
We thank R. Valenti, and P. Hirschfeld for discussions. F.A. and I.E. were supported by the joint
DFG-ANR Project (ER 463/8-1).

\bibliographystyle{apsrev4-1}
\bibliography{paper1144bib}

\begin{thebibliography}{31}%
\makeatletter
\providecommand \@ifxundefined [1]{%
 \@ifx{#1\undefined}
}%
\providecommand \@ifnum [1]{%
 \ifnum #1\expandafter \@firstoftwo
 \else \expandafter \@secondoftwo
 \fi
}%
\providecommand \@ifx [1]{%
 \ifx #1\expandafter \@firstoftwo
 \else \expandafter \@secondoftwo
 \fi
}%
\providecommand \natexlab [1]{#1}%
\providecommand \enquote  [1]{``#1''}%
\providecommand \bibnamefont  [1]{#1}%
\providecommand \bibfnamefont [1]{#1}%
\providecommand \citenamefont [1]{#1}%
\providecommand \href@noop [0]{\@secondoftwo}%
\providecommand \href [0]{\begingroup \@sanitize@url \@href}%
\providecommand \@href[1]{\@@startlink{#1}\@@href}%
\providecommand \@@href[1]{\endgroup#1\@@endlink}%
\providecommand \@sanitize@url [0]{\catcode `\\12\catcode `\$12\catcode
  `\&12\catcode `\#12\catcode `\^12\catcode `\_12\catcode `\%12\relax}%
\providecommand \@@startlink[1]{}%
\providecommand \@@endlink[0]{}%
\providecommand \url  [0]{\begingroup\@sanitize@url \@url }%
\providecommand \@url [1]{\endgroup\@href {#1}{\urlprefix }}%
\providecommand \urlprefix  [0]{URL }%
\providecommand \Eprint [0]{\href }%
\providecommand \doibase [0]{http://dx.doi.org/}%
\providecommand \selectlanguage [0]{\@gobble}%
\providecommand \bibinfo  [0]{\@secondoftwo}%
\providecommand \bibfield  [0]{\@secondoftwo}%
\providecommand \translation [1]{[#1]}%
\providecommand \BibitemOpen [0]{}%
\providecommand \bibitemStop [0]{}%
\providecommand \bibitemNoStop [0]{.\EOS\space}%
\providecommand \EOS [0]{\spacefactor3000\relax}%
\providecommand \BibitemShut  [1]{\csname bibitem#1\endcsname}%
\let\auto@bib@innerbib\@empty
\bibitem [{\citenamefont {Kamihara}\ \emph {et~al.}(2008)\citenamefont
  {Kamihara}, \citenamefont {Watanabe}, \citenamefont {Hirano},\ and\
  \citenamefont {Hosono}}]{Kamihara.JACS.130.3296(2008)}%
  \BibitemOpen
  \bibfield  {author} {\bibinfo {author} {\bibfnamefont {Y.}~\bibnamefont
  {Kamihara}}, \bibinfo {author} {\bibfnamefont {T.}~\bibnamefont {Watanabe}},
  \bibinfo {author} {\bibfnamefont {M.}~\bibnamefont {Hirano}}, \ and\ \bibinfo
  {author} {\bibfnamefont {H.}~\bibnamefont {Hosono}},\ }\href {\doibase
  10.1021/ja800073m} {\bibfield  {journal} {\bibinfo  {journal} {J. Am. Chem.
  Soc.}\ }\textbf {\bibinfo {volume} {130}},\ \bibinfo {pages} {3296} (\bibinfo
  {year} {2008})}\BibitemShut {NoStop}%
\bibitem [{\citenamefont {Johnston}(2010)}]{Johnston_review}%
  \BibitemOpen
  \bibfield  {author} {\bibinfo {author} {\bibfnamefont {D.~C.}\ \bibnamefont
  {Johnston}},\ }\href {\doibase 10.1080/00018732.2010.513480} {\bibfield
  {journal} {\bibinfo  {journal} {Adv. Phys.}\ }\textbf {\bibinfo {volume}
  {59}},\ \bibinfo {pages} {503} (\bibinfo {year} {2010})}\BibitemShut
  {NoStop}%
\bibitem [{\citenamefont {Hirschfeld}(2016)}]{Hirschfeld_review}%
  \BibitemOpen
  \bibfield  {author} {\bibinfo {author} {\bibfnamefont {P.}~\bibnamefont
  {Hirschfeld}},\ }\href {\doibase 10.1016/j.crhy.2015.10.002} {\bibfield
  {journal} {\bibinfo  {journal} {C. R. Phys.}\ }\textbf {\bibinfo {volume}
  {17}},\ \bibinfo {pages} {197} (\bibinfo {year} {2016})}\BibitemShut
  {NoStop}%
\bibitem [{\citenamefont {Chubukov}(2012)}]{Chubukov_review}%
  \BibitemOpen
  \bibfield  {author} {\bibinfo {author} {\bibfnamefont {A.~V.}\ \bibnamefont
  {Chubukov}},\ }\href {\doibase 10.1146/annurev-conmatphys-020911-125055}
  {\bibfield  {journal} {\bibinfo  {journal} {Annu. Rev. Condens. Matter
  Phys.}\ }\textbf {\bibinfo {volume} {3}},\ \bibinfo {pages} {57} (\bibinfo
  {year} {2012})}\BibitemShut {NoStop}%
\bibitem [{\citenamefont {Iyo}\ \emph {et~al.}(2016)\citenamefont {Iyo},
  \citenamefont {Kawashima}, \citenamefont {Kinjo}, \citenamefont {Nishio},
  \citenamefont {Ishida}, \citenamefont {Fujihisa}, \citenamefont {Gotoh},
  \citenamefont {Kihou}, \citenamefont {Eisaki},\ and\ \citenamefont
  {Yoshida}}]{Iyo.JACS.138.3410(2016)}%
  \BibitemOpen
  \bibfield  {author} {\bibinfo {author} {\bibfnamefont {A.}~\bibnamefont
  {Iyo}}, \bibinfo {author} {\bibfnamefont {K.}~\bibnamefont {Kawashima}},
  \bibinfo {author} {\bibfnamefont {T.}~\bibnamefont {Kinjo}}, \bibinfo
  {author} {\bibfnamefont {T.}~\bibnamefont {Nishio}}, \bibinfo {author}
  {\bibfnamefont {S.}~\bibnamefont {Ishida}}, \bibinfo {author} {\bibfnamefont
  {H.}~\bibnamefont {Fujihisa}}, \bibinfo {author} {\bibfnamefont
  {Y.}~\bibnamefont {Gotoh}}, \bibinfo {author} {\bibfnamefont
  {K.}~\bibnamefont {Kihou}}, \bibinfo {author} {\bibfnamefont
  {H.}~\bibnamefont {Eisaki}}, \ and\ \bibinfo {author} {\bibfnamefont
  {Y.}~\bibnamefont {Yoshida}},\ }\href {\doibase 10.1021/jacs.5b12571}
  {\bibfield  {journal} {\bibinfo  {journal} {J. Am. Chem. Soc.}\ }\textbf
  {\bibinfo {volume} {138}},\ \bibinfo {pages} {3410} (\bibinfo {year}
  {2016})}\BibitemShut {NoStop}%
\bibitem [{\citenamefont {Mou}\ \emph {et~al.}(2016)\citenamefont {Mou},
  \citenamefont {Kong}, \citenamefont {Meier}, \citenamefont {Lochner},
  \citenamefont {Wang}, \citenamefont {Lin}, \citenamefont {Wu}, \citenamefont
  {Bud'ko}, \citenamefont {Eremin}, \citenamefont {Johnson}, \citenamefont
  {Canfield},\ and\ \citenamefont {Kaminski}}]{Mou.PRL.117.277001(2016)}%
  \BibitemOpen
  \bibfield  {author} {\bibinfo {author} {\bibfnamefont {D.}~\bibnamefont
  {Mou}}, \bibinfo {author} {\bibfnamefont {T.}~\bibnamefont {Kong}}, \bibinfo
  {author} {\bibfnamefont {W.~R.}\ \bibnamefont {Meier}}, \bibinfo {author}
  {\bibfnamefont {F.}~\bibnamefont {Lochner}}, \bibinfo {author} {\bibfnamefont
  {L.-L.}\ \bibnamefont {Wang}}, \bibinfo {author} {\bibfnamefont
  {Q.}~\bibnamefont {Lin}}, \bibinfo {author} {\bibfnamefont {Y.}~\bibnamefont
  {Wu}}, \bibinfo {author} {\bibfnamefont {S.~L.}\ \bibnamefont {Bud'ko}},
  \bibinfo {author} {\bibfnamefont {I.}~\bibnamefont {Eremin}}, \bibinfo
  {author} {\bibfnamefont {D.~D.}\ \bibnamefont {Johnson}}, \bibinfo {author}
  {\bibfnamefont {P.~C.}\ \bibnamefont {Canfield}}, \ and\ \bibinfo {author}
  {\bibfnamefont {A.}~\bibnamefont {Kaminski}},\ }\href {\doibase
  10.1103/PhysRevLett.117.277001} {\bibfield  {journal} {\bibinfo  {journal}
  {Phys. Rev. Lett.}\ }\textbf {\bibinfo {volume} {117}},\ \bibinfo {pages}
  {277001} (\bibinfo {year} {2016})}\BibitemShut {NoStop}%
\bibitem [{\citenamefont {Yang}\ \emph {et~al.}(2017)\citenamefont {Yang},
  \citenamefont {Dai}, \citenamefont {Xu}, \citenamefont {Zhang}, \citenamefont
  {Qiu}, \citenamefont {Sui}, \citenamefont {Homes},\ and\ \citenamefont
  {Qiu}}]{Yang.PRB.95.064506(2017)}%
  \BibitemOpen
  \bibfield  {author} {\bibinfo {author} {\bibfnamefont {R.}~\bibnamefont
  {Yang}}, \bibinfo {author} {\bibfnamefont {Y.}~\bibnamefont {Dai}}, \bibinfo
  {author} {\bibfnamefont {B.}~\bibnamefont {Xu}}, \bibinfo {author}
  {\bibfnamefont {W.}~\bibnamefont {Zhang}}, \bibinfo {author} {\bibfnamefont
  {Z.}~\bibnamefont {Qiu}}, \bibinfo {author} {\bibfnamefont {Q.}~\bibnamefont
  {Sui}}, \bibinfo {author} {\bibfnamefont {C.~C.}\ \bibnamefont {Homes}}, \
  and\ \bibinfo {author} {\bibfnamefont {X.}~\bibnamefont {Qiu}},\ }\href
  {\doibase 10.1103/PhysRevB.95.064506} {\bibfield  {journal} {\bibinfo
  {journal} {Phys. Rev. B}\ }\textbf {\bibinfo {volume} {95}},\ \bibinfo
  {pages} {064506} (\bibinfo {year} {2017})}\BibitemShut {NoStop}%
\bibitem [{\citenamefont {Cho}\ \emph {et~al.}(2017)\citenamefont {Cho},
  \citenamefont {Fente}, \citenamefont {Teknowijoyo}, \citenamefont {Tanatar},
  \citenamefont {Joshi}, \citenamefont {Nusran}, \citenamefont {Kong},
  \citenamefont {Meier}, \citenamefont {Kaluarachchi}, \citenamefont
  {Guillam{\'o}n}, \citenamefont {Suderow}, \citenamefont {Bud'ko},
  \citenamefont {Canfield},\ and\ \citenamefont
  {Prozorov}}]{Cho.PRB.95.100502(2017)}%
  \BibitemOpen
  \bibfield  {author} {\bibinfo {author} {\bibfnamefont {K.}~\bibnamefont
  {Cho}}, \bibinfo {author} {\bibfnamefont {A.}~\bibnamefont {Fente}}, \bibinfo
  {author} {\bibfnamefont {S.}~\bibnamefont {Teknowijoyo}}, \bibinfo {author}
  {\bibfnamefont {M.~A.}\ \bibnamefont {Tanatar}}, \bibinfo {author}
  {\bibfnamefont {K.~R.}\ \bibnamefont {Joshi}}, \bibinfo {author}
  {\bibfnamefont {N.~M.}\ \bibnamefont {Nusran}}, \bibinfo {author}
  {\bibfnamefont {T.}~\bibnamefont {Kong}}, \bibinfo {author} {\bibfnamefont
  {W.~R.}\ \bibnamefont {Meier}}, \bibinfo {author} {\bibfnamefont
  {U.}~\bibnamefont {Kaluarachchi}}, \bibinfo {author} {\bibfnamefont
  {I.}~\bibnamefont {Guillam{\'o}n}}, \bibinfo {author} {\bibfnamefont
  {H.}~\bibnamefont {Suderow}}, \bibinfo {author} {\bibfnamefont {S.~L.}\
  \bibnamefont {Bud'ko}}, \bibinfo {author} {\bibfnamefont {P.~C.}\
  \bibnamefont {Canfield}}, \ and\ \bibinfo {author} {\bibfnamefont
  {R.}~\bibnamefont {Prozorov}},\ }\href {\doibase 10.1103/PhysRevB.95.100502}
  {\bibfield  {journal} {\bibinfo  {journal} {Phys. Rev. B}\ }\textbf {\bibinfo
  {volume} {95}},\ \bibinfo {pages} {100502(R)} (\bibinfo {year}
  {2017})}\BibitemShut {NoStop}%
\bibitem [{\citenamefont {Shi}\ and\ \citenamefont
  {Wang}(2016)}]{Shi.JPSJ.85.124714(2016)}%
  \BibitemOpen
  \bibfield  {author} {\bibinfo {author} {\bibfnamefont {X.}~\bibnamefont
  {Shi}}\ and\ \bibinfo {author} {\bibfnamefont {G.}~\bibnamefont {Wang}},\
  }\href {\doibase 10.7566/JPSJ.85.124714} {\bibfield  {journal} {\bibinfo
  {journal} {J. Phys. Soc. Jpn.}\ }\textbf {\bibinfo {volume} {85}},\ \bibinfo
  {pages} {124714} (\bibinfo {year} {2016})}\BibitemShut {NoStop}%
\bibitem [{\citenamefont {Meier}\ \emph {et~al.}(2016)\citenamefont {Meier},
  \citenamefont {Kong}, \citenamefont {Kaluarachchi}, \citenamefont {Jo},
  \citenamefont {Drachuck}, \citenamefont {B\"ohmer}, \citenamefont {Saunders},
  \citenamefont {Sapkota}, \citenamefont {Kreyssig}, \citenamefont {Tanatar},
  \citenamefont {Prozorov}, \citenamefont {Goldman}, \citenamefont {Balakirev},
  \citenamefont {Gurevich}, \citenamefont {Bud’ko},\ and\ \citenamefont
  {Canfield}}]{Meier2016}%
  \BibitemOpen
  \bibfield  {author} {\bibinfo {author} {\bibfnamefont {W.}~\bibnamefont
  {Meier}}, \bibinfo {author} {\bibfnamefont {T.}~\bibnamefont {Kong}},
  \bibinfo {author} {\bibfnamefont {V.}~\bibnamefont {Kaluarachchi},
  \bibfnamefont {U.~S.and~Taufour}}, \bibinfo {author} {\bibfnamefont {N.~H.}\
  \bibnamefont {Jo}}, \bibinfo {author} {\bibfnamefont {G.}~\bibnamefont
  {Drachuck}}, \bibinfo {author} {\bibfnamefont {A.}~\bibnamefont {B\"ohmer}},
  \bibinfo {author} {\bibfnamefont {S.}~\bibnamefont {Saunders}}, \bibinfo
  {author} {\bibfnamefont {A.}~\bibnamefont {Sapkota}}, \bibinfo {author}
  {\bibfnamefont {A.}~\bibnamefont {Kreyssig}}, \bibinfo {author}
  {\bibfnamefont {M.}~\bibnamefont {Tanatar}}, \bibinfo {author} {\bibfnamefont
  {R.}~\bibnamefont {Prozorov}}, \bibinfo {author} {\bibfnamefont
  {A.}~\bibnamefont {Goldman}}, \bibinfo {author} {\bibfnamefont
  {F.}~\bibnamefont {Balakirev}}, \bibinfo {author} {\bibfnamefont
  {A.}~\bibnamefont {Gurevich}}, \bibinfo {author} {\bibfnamefont
  {S.}~\bibnamefont {Bud’ko}}, \ and\ \bibinfo {author} {\bibfnamefont
  {P.}~\bibnamefont {Canfield}},\ }\href@noop {} {\bibfield  {journal}
  {\bibinfo  {journal} {Phys. Rev. B}\ }\textbf {\bibinfo {volume} {94}},\
  \bibinfo {pages} {064501} (\bibinfo {year} {2016})}\BibitemShut {NoStop}%
\bibitem [{\citenamefont {Liu}\ \emph {et~al.}(2014)\citenamefont {Liu},
  \citenamefont {Tanatar}, \citenamefont {Straszheim}, \citenamefont {Jensen},
  \citenamefont {Dennis}, \citenamefont {McCallum}, \citenamefont {Kogan},
  \citenamefont {Prozorov},\ and\ \citenamefont
  {Lograsso}}]{Liu.PRB.89.134504(2014)}%
  \BibitemOpen
  \bibfield  {author} {\bibinfo {author} {\bibfnamefont {Y.}~\bibnamefont
  {Liu}}, \bibinfo {author} {\bibfnamefont {M.~A.}\ \bibnamefont {Tanatar}},
  \bibinfo {author} {\bibfnamefont {W.~E.}\ \bibnamefont {Straszheim}},
  \bibinfo {author} {\bibfnamefont {B.}~\bibnamefont {Jensen}}, \bibinfo
  {author} {\bibfnamefont {K.~W.}\ \bibnamefont {Dennis}}, \bibinfo {author}
  {\bibfnamefont {R.~W.}\ \bibnamefont {McCallum}}, \bibinfo {author}
  {\bibfnamefont {V.~G.}\ \bibnamefont {Kogan}}, \bibinfo {author}
  {\bibfnamefont {R.}~\bibnamefont {Prozorov}}, \ and\ \bibinfo {author}
  {\bibfnamefont {T.~A.}\ \bibnamefont {Lograsso}},\ }\href {\doibase
  10.1103/PhysRevB.89.134504} {\bibfield  {journal} {\bibinfo  {journal} {Phys.
  Rev. B}\ }\textbf {\bibinfo {volume} {89}},\ \bibinfo {pages} {134504}
  (\bibinfo {year} {2014})}\BibitemShut {NoStop}%
\bibitem [{\citenamefont {Cho}\ \emph {et~al.}(2016)\citenamefont {Cho},
  \citenamefont {Ko{\'n}czykowski}, \citenamefont {Teknowijoyo}, \citenamefont
  {Tanatar}, \citenamefont {Liu}, \citenamefont {Lograsso}, \citenamefont
  {Straszheim}, \citenamefont {Mishra}, \citenamefont {Maiti}, \citenamefont
  {Hirschfeld},\ and\ \citenamefont {Prozorov}}]{Cho2016}%
  \BibitemOpen
  \bibfield  {author} {\bibinfo {author} {\bibfnamefont {K.}~\bibnamefont
  {Cho}}, \bibinfo {author} {\bibfnamefont {M.}~\bibnamefont
  {Ko{\'n}czykowski}}, \bibinfo {author} {\bibfnamefont {S.}~\bibnamefont
  {Teknowijoyo}}, \bibinfo {author} {\bibfnamefont {M.~A.}\ \bibnamefont
  {Tanatar}}, \bibinfo {author} {\bibfnamefont {Y.}~\bibnamefont {Liu}},
  \bibinfo {author} {\bibfnamefont {T.~A.}\ \bibnamefont {Lograsso}}, \bibinfo
  {author} {\bibfnamefont {W.~E.}\ \bibnamefont {Straszheim}}, \bibinfo
  {author} {\bibfnamefont {V.}~\bibnamefont {Mishra}}, \bibinfo {author}
  {\bibfnamefont {S.}~\bibnamefont {Maiti}}, \bibinfo {author} {\bibfnamefont
  {P.~J.}\ \bibnamefont {Hirschfeld}}, \ and\ \bibinfo {author} {\bibfnamefont
  {R.}~\bibnamefont {Prozorov}},\ }\href {\doibase 10.1126/sciadv.1600807}
  {\bibfield  {journal} {\bibinfo  {journal} {Science Advances}\ }\textbf
  {\bibinfo {volume} {2}},\ \bibinfo {pages} {e1600807} (\bibinfo {year}
  {2016})}\BibitemShut {NoStop}%
\bibitem [{\citenamefont {Maiti}\ \emph
  {et~al.}(2011{\natexlab{a}})\citenamefont {Maiti}, \citenamefont {Korshunov},
  \citenamefont {Maier}, \citenamefont {Hirschfeld},\ and\ \citenamefont
  {Chubukov}}]{Maiti2011a}%
  \BibitemOpen
  \bibfield  {author} {\bibinfo {author} {\bibfnamefont {S.}~\bibnamefont
  {Maiti}}, \bibinfo {author} {\bibfnamefont {M.~M.}\ \bibnamefont
  {Korshunov}}, \bibinfo {author} {\bibfnamefont {T.~A.}\ \bibnamefont
  {Maier}}, \bibinfo {author} {\bibfnamefont {P.~J.}\ \bibnamefont
  {Hirschfeld}}, \ and\ \bibinfo {author} {\bibfnamefont {A.~V.}\ \bibnamefont
  {Chubukov}},\ }\href {\doibase 10.1103/PhysRevLett.107.147002} {\bibfield
  {journal} {\bibinfo  {journal} {Phys. Rev. Lett.}\ }\textbf {\bibinfo
  {volume} {107}},\ \bibinfo {pages} {147002} (\bibinfo {year}
  {2011}{\natexlab{a}})}\BibitemShut {NoStop}%
\bibitem [{\citenamefont {Maiti}\ \emph
  {et~al.}(2011{\natexlab{b}})\citenamefont {Maiti}, \citenamefont {Korshunov},
  \citenamefont {Maier}, \citenamefont {Hirschfeld},\ and\ \citenamefont
  {Chubukov}}]{Maiti2011b}%
  \BibitemOpen
  \bibfield  {author} {\bibinfo {author} {\bibfnamefont {S.}~\bibnamefont
  {Maiti}}, \bibinfo {author} {\bibfnamefont {M.~M.}\ \bibnamefont
  {Korshunov}}, \bibinfo {author} {\bibfnamefont {T.~A.}\ \bibnamefont
  {Maier}}, \bibinfo {author} {\bibfnamefont {P.~J.}\ \bibnamefont
  {Hirschfeld}}, \ and\ \bibinfo {author} {\bibfnamefont {A.~V.}\ \bibnamefont
  {Chubukov}},\ }\href {\doibase 10.1103/PhysRevB.84.224505} {\bibfield
  {journal} {\bibinfo  {journal} {Phys. Rev. B}\ }\textbf {\bibinfo {volume}
  {84}},\ \bibinfo {pages} {224505} (\bibinfo {year}
  {2011}{\natexlab{b}})}\BibitemShut {NoStop}%
\bibitem [{\citenamefont {Ahn}\ \emph {et~al.}(2014)\citenamefont {Ahn},
  \citenamefont {Eremin}, \citenamefont {Knolle}, \citenamefont {Zabolotnyy},
  \citenamefont {Borisenko}, \citenamefont {B{\"u}chner},\ and\ \citenamefont
  {Chubukov}}]{Ahn.PRB.89.144513(2014)}%
  \BibitemOpen
  \bibfield  {author} {\bibinfo {author} {\bibfnamefont {F.}~\bibnamefont
  {Ahn}}, \bibinfo {author} {\bibfnamefont {I.}~\bibnamefont {Eremin}},
  \bibinfo {author} {\bibfnamefont {J.}~\bibnamefont {Knolle}}, \bibinfo
  {author} {\bibfnamefont {V.~B.}\ \bibnamefont {Zabolotnyy}}, \bibinfo
  {author} {\bibfnamefont {S.~V.}\ \bibnamefont {Borisenko}}, \bibinfo {author}
  {\bibfnamefont {B.}~\bibnamefont {B{\"u}chner}}, \ and\ \bibinfo {author}
  {\bibfnamefont {A.~V.}\ \bibnamefont {Chubukov}},\ }\href {\doibase
  10.1103/PhysRevB.89.144513} {\bibfield  {journal} {\bibinfo  {journal} {Phys.
  Rev. B}\ }\textbf {\bibinfo {volume} {89}},\ \bibinfo {pages} {144513}
  (\bibinfo {year} {2014})}\BibitemShut {NoStop}%
\bibitem [{\citenamefont {Maiti}\ \emph {et~al.}(2012)\citenamefont {Maiti},
  \citenamefont {Korshunov},\ and\ \citenamefont
  {Chubukov}}]{Maiti.PRB.85.014511(2012)}%
  \BibitemOpen
  \bibfield  {author} {\bibinfo {author} {\bibfnamefont {S.}~\bibnamefont
  {Maiti}}, \bibinfo {author} {\bibfnamefont {M.~M.}\ \bibnamefont
  {Korshunov}}, \ and\ \bibinfo {author} {\bibfnamefont {A.~V.}\ \bibnamefont
  {Chubukov}},\ }\href {\doibase 10.1103/PhysRevB.85.014511} {\bibfield
  {journal} {\bibinfo  {journal} {Phys. Rev. B}\ }\textbf {\bibinfo {volume}
  {85}},\ \bibinfo {pages} {014511} (\bibinfo {year} {2012})}\BibitemShut
  {NoStop}%
\bibitem [{\citenamefont {Kresse}\ and\ \citenamefont
  {Hafner}(1993)}]{Kresse.PRB.47.558(1993)}%
  \BibitemOpen
  \bibfield  {author} {\bibinfo {author} {\bibfnamefont {G.}~\bibnamefont
  {Kresse}}\ and\ \bibinfo {author} {\bibfnamefont {J.}~\bibnamefont
  {Hafner}},\ }\href {\doibase 10.1103/PhysRevB.47.558} {\bibfield  {journal}
  {\bibinfo  {journal} {Phys. Rev. B}\ }\textbf {\bibinfo {volume} {47}},\
  \bibinfo {pages} {558} (\bibinfo {year} {1993})}\BibitemShut {NoStop}%
\bibitem [{\citenamefont {Kresse}\ and\ \citenamefont
  {Furthm{\"u}ller}(1996{\natexlab{a}})}]{Kresse.ComMatSci.6.15(1996)}%
  \BibitemOpen
  \bibfield  {author} {\bibinfo {author} {\bibfnamefont {G.}~\bibnamefont
  {Kresse}}\ and\ \bibinfo {author} {\bibfnamefont {J.}~\bibnamefont
  {Furthm{\"u}ller}},\ }\href {\doibase 10.1016/0927-0256(96)00008-0}
  {\bibfield  {journal} {\bibinfo  {journal} {Comput. Mater. Sci.}\ }\textbf
  {\bibinfo {volume} {6}},\ \bibinfo {pages} {15} (\bibinfo {year}
  {1996}{\natexlab{a}})}\BibitemShut {NoStop}%
\bibitem [{\citenamefont {Kresse}\ and\ \citenamefont
  {Furthm{\"u}ller}(1996{\natexlab{b}})}]{Kresse.PRB.54.11169(1996)}%
  \BibitemOpen
  \bibfield  {author} {\bibinfo {author} {\bibfnamefont {G.}~\bibnamefont
  {Kresse}}\ and\ \bibinfo {author} {\bibfnamefont {J.}~\bibnamefont
  {Furthm{\"u}ller}},\ }\href {\doibase 10.1103/PhysRevB.54.11169} {\bibfield
  {journal} {\bibinfo  {journal} {Phys. Rev. B}\ }\textbf {\bibinfo {volume}
  {54}},\ \bibinfo {pages} {11169} (\bibinfo {year}
  {1996}{\natexlab{b}})}\BibitemShut {NoStop}%
\bibitem [{\citenamefont {Bl{\"o}chl}(1994)}]{Blochl.PRB.50.17953(1994)}%
  \BibitemOpen
  \bibfield  {author} {\bibinfo {author} {\bibfnamefont {P.~E.}\ \bibnamefont
  {Bl{\"o}chl}},\ }\href {\doibase 10.1103/PhysRevB.50.17953} {\bibfield
  {journal} {\bibinfo  {journal} {Phys. Rev. B}\ }\textbf {\bibinfo {volume}
  {50}},\ \bibinfo {pages} {17953} (\bibinfo {year} {1994})}\BibitemShut
  {NoStop}%
\bibitem [{\citenamefont {R{\'o}zsa}\ and\ \citenamefont
  {Schuster}(1981)}]{Rozsa.ZNB.36.1668(1981)}%
  \BibitemOpen
  \bibfield  {author} {\bibinfo {author} {\bibfnamefont {S.}~\bibnamefont
  {R{\'o}zsa}}\ and\ \bibinfo {author} {\bibfnamefont {H.-U.}\ \bibnamefont
  {Schuster}},\ }\href {http://www.znaturforsch.com/ab/v36b/36b1668.pdf}
  {\bibfield  {journal} {\bibinfo  {journal} {Z. Naturforsch. B}\ }\textbf
  {\bibinfo {volume} {36}},\ \bibinfo {pages} {1668} (\bibinfo {year}
  {1981})}\BibitemShut {NoStop}%
\bibitem [{\citenamefont {Kreyssig}\ \emph {et~al.}(2008)\citenamefont
  {Kreyssig}, \citenamefont {Green}, \citenamefont {Lee}, \citenamefont
  {Samolyuk}, \citenamefont {Zajdel}, \citenamefont {Lynn}, \citenamefont
  {Bud'ko}, \citenamefont {Torikachvili}, \citenamefont {Ni}, \citenamefont
  {Nandi}, \citenamefont {Le{\~a}o}, \citenamefont {Poulton}, \citenamefont
  {Argyriou}, \citenamefont {Harmon}, \citenamefont {McQueeney}, \citenamefont
  {Canfield},\ and\ \citenamefont {Goldman}}]{Kreyssig.PRB.78.184517(2008)}%
  \BibitemOpen
  \bibfield  {author} {\bibinfo {author} {\bibfnamefont {A.}~\bibnamefont
  {Kreyssig}}, \bibinfo {author} {\bibfnamefont {M.~A.}\ \bibnamefont {Green}},
  \bibinfo {author} {\bibfnamefont {Y.}~\bibnamefont {Lee}}, \bibinfo {author}
  {\bibfnamefont {G.~D.}\ \bibnamefont {Samolyuk}}, \bibinfo {author}
  {\bibfnamefont {P.}~\bibnamefont {Zajdel}}, \bibinfo {author} {\bibfnamefont
  {J.~W.}\ \bibnamefont {Lynn}}, \bibinfo {author} {\bibfnamefont {S.~L.}\
  \bibnamefont {Bud'ko}}, \bibinfo {author} {\bibfnamefont {M.~S.}\
  \bibnamefont {Torikachvili}}, \bibinfo {author} {\bibfnamefont
  {N.}~\bibnamefont {Ni}}, \bibinfo {author} {\bibfnamefont {S.}~\bibnamefont
  {Nandi}}, \bibinfo {author} {\bibfnamefont {J.~B.}\ \bibnamefont {Le{\~a}o}},
  \bibinfo {author} {\bibfnamefont {S.~J.}\ \bibnamefont {Poulton}}, \bibinfo
  {author} {\bibfnamefont {D.~N.}\ \bibnamefont {Argyriou}}, \bibinfo {author}
  {\bibfnamefont {B.~N.}\ \bibnamefont {Harmon}}, \bibinfo {author}
  {\bibfnamefont {R.~J.}\ \bibnamefont {McQueeney}}, \bibinfo {author}
  {\bibfnamefont {P.~C.}\ \bibnamefont {Canfield}}, \ and\ \bibinfo {author}
  {\bibfnamefont {A.~I.}\ \bibnamefont {Goldman}},\ }\href {\doibase
  10.1103/PhysRevB.78.184517} {\bibfield  {journal} {\bibinfo  {journal} {Phys.
  Rev. B}\ }\textbf {\bibinfo {volume} {78}},\ \bibinfo {pages} {184517}
  (\bibinfo {year} {2008})}\BibitemShut {NoStop}%
\bibitem [{\citenamefont {Perdew}\ \emph {et~al.}(1996)\citenamefont {Perdew},
  \citenamefont {Burke},\ and\ \citenamefont
  {Ernzerhof}}]{Perdew.PRL.77.3865(1996)}%
  \BibitemOpen
  \bibfield  {author} {\bibinfo {author} {\bibfnamefont {J.~P.}\ \bibnamefont
  {Perdew}}, \bibinfo {author} {\bibfnamefont {K.}~\bibnamefont {Burke}}, \
  and\ \bibinfo {author} {\bibfnamefont {M.}~\bibnamefont {Ernzerhof}},\ }\href
  {\doibase 10.1103/PhysRevLett.77.3865} {\bibfield  {journal} {\bibinfo
  {journal} {Phys. Rev. Lett.}\ }\textbf {\bibinfo {volume} {77}},\ \bibinfo
  {pages} {3865} (\bibinfo {year} {1996})}\BibitemShut {NoStop}%
\bibitem [{\citenamefont {Mostofi}\ \emph {et~al.}(2014)\citenamefont
  {Mostofi}, \citenamefont {Yates}, \citenamefont {Pizzi}, \citenamefont {Lee},
  \citenamefont {Souza}, \citenamefont {Vanderbilt},\ and\ \citenamefont
  {Marzari}}]{Mostofi.CompPhysCom.185.2309(2014)}%
  \BibitemOpen
  \bibfield  {author} {\bibinfo {author} {\bibfnamefont {A.~A.}\ \bibnamefont
  {Mostofi}}, \bibinfo {author} {\bibfnamefont {J.~R.}\ \bibnamefont {Yates}},
  \bibinfo {author} {\bibfnamefont {G.}~\bibnamefont {Pizzi}}, \bibinfo
  {author} {\bibfnamefont {Y.-S.}\ \bibnamefont {Lee}}, \bibinfo {author}
  {\bibfnamefont {I.}~\bibnamefont {Souza}}, \bibinfo {author} {\bibfnamefont
  {D.}~\bibnamefont {Vanderbilt}}, \ and\ \bibinfo {author} {\bibfnamefont
  {N.}~\bibnamefont {Marzari}},\ }\href {\doibase 10.1016/j.cpc.2014.05.003}
  {\bibfield  {journal} {\bibinfo  {journal} {Comput. Phys. Commun.}\ }\textbf
  {\bibinfo {volume} {185}},\ \bibinfo {pages} {2309} (\bibinfo {year}
  {2014})}\BibitemShut {NoStop}%
\bibitem [{\citenamefont {Marzari}\ and\ \citenamefont
  {Vanderbilt}(1997)}]{Marzari.PRB.56.12847(1997)}%
  \BibitemOpen
  \bibfield  {author} {\bibinfo {author} {\bibfnamefont {N.}~\bibnamefont
  {Marzari}}\ and\ \bibinfo {author} {\bibfnamefont {D.}~\bibnamefont
  {Vanderbilt}},\ }\href {\doibase 10.1103/PhysRevB.56.12847} {\bibfield
  {journal} {\bibinfo  {journal} {Phys. Rev. B}\ }\textbf {\bibinfo {volume}
  {56}},\ \bibinfo {pages} {12847} (\bibinfo {year} {1997})}\BibitemShut
  {NoStop}%
\bibitem [{\citenamefont {Meier}\ \emph {et~al.}(2017)\citenamefont {Meier},
  \citenamefont {Ding}, \citenamefont {Kreyssig}, \citenamefont {Bud'ko},
  \citenamefont {Sapkota}, \citenamefont {Kothapalli}, \citenamefont {Borisov},
  \citenamefont {Valentí}, \citenamefont {Batista}, \citenamefont {Orth},
  \citenamefont {Fernandes}, \citenamefont {Goldman}, \citenamefont {Furukawa},
  \citenamefont {B\"ohmer},\ and\ \citenamefont {Canfield}}]{Meier2017}%
  \BibitemOpen
  \bibfield  {author} {\bibinfo {author} {\bibfnamefont {W.}~\bibnamefont
  {Meier}}, \bibinfo {author} {\bibfnamefont {Q.-P.}\ \bibnamefont {Ding}},
  \bibinfo {author} {\bibfnamefont {A.}~\bibnamefont {Kreyssig}}, \bibinfo
  {author} {\bibfnamefont {S.~L.}\ \bibnamefont {Bud'ko}}, \bibinfo {author}
  {\bibfnamefont {A.}~\bibnamefont {Sapkota}}, \bibinfo {author} {\bibfnamefont
  {K.}~\bibnamefont {Kothapalli}}, \bibinfo {author} {\bibfnamefont
  {V.}~\bibnamefont {Borisov}}, \bibinfo {author} {\bibfnamefont
  {R.}~\bibnamefont {Valentí}}, \bibinfo {author} {\bibfnamefont {C.~D.}\
  \bibnamefont {Batista}}, \bibinfo {author} {\bibfnamefont {P.~P.}\
  \bibnamefont {Orth}}, \bibinfo {author} {\bibfnamefont {R.~M.}\ \bibnamefont
  {Fernandes}}, \bibinfo {author} {\bibfnamefont {A.~I.}\ \bibnamefont
  {Goldman}}, \bibinfo {author} {\bibfnamefont {Y.}~\bibnamefont {Furukawa}},
  \bibinfo {author} {\bibfnamefont {A.~E.}\ \bibnamefont {B\"ohmer}}, \ and\
  \bibinfo {author} {\bibfnamefont {P.~C.}\ \bibnamefont {Canfield}},\
  }\href@noop {} {\bibfield  {journal} {\bibinfo  {journal} {arXiv:1706.01067
  (unpublished)}\ } (\bibinfo {year} {2017})}\BibitemShut {NoStop}%
\bibitem [{\citenamefont {Eschrig}\ and\ \citenamefont
  {Koepernik}(2009)}]{Eschrig.PRB.80.104503(2009)}%
  \BibitemOpen
  \bibfield  {author} {\bibinfo {author} {\bibfnamefont {H.}~\bibnamefont
  {Eschrig}}\ and\ \bibinfo {author} {\bibfnamefont {K.}~\bibnamefont
  {Koepernik}},\ }\href {\doibase 10.1103/PhysRevB.80.104503} {\bibfield
  {journal} {\bibinfo  {journal} {Phys. Rev. B}\ }\textbf {\bibinfo {volume}
  {80}},\ \bibinfo {pages} {104503} (\bibinfo {year} {2009})}\BibitemShut
  {NoStop}%
\bibitem [{\citenamefont {Graser}\ \emph {et~al.}(2009)\citenamefont {Graser},
  \citenamefont {Maier}, \citenamefont {Hirschfeld},\ and\ \citenamefont
  {Scalapino}}]{Graser.NJP.11.025016(2009)}%
  \BibitemOpen
  \bibfield  {author} {\bibinfo {author} {\bibfnamefont {S.}~\bibnamefont
  {Graser}}, \bibinfo {author} {\bibfnamefont {T.~A.}\ \bibnamefont {Maier}},
  \bibinfo {author} {\bibfnamefont {P.~J.}\ \bibnamefont {Hirschfeld}}, \ and\
  \bibinfo {author} {\bibfnamefont {D.~J.}\ \bibnamefont {Scalapino}},\ }\href
  {\doibase 10.1088/1367-2630/11/2/025016} {\bibfield  {journal} {\bibinfo
  {journal} {New J. Phys.}\ }\textbf {\bibinfo {volume} {11}},\ \bibinfo
  {pages} {025016} (\bibinfo {year} {2009})}\BibitemShut {NoStop}%
\bibitem [{\citenamefont {Ikeda}\ \emph {et~al.}(2010)\citenamefont {Ikeda},
  \citenamefont {Arita},\ and\ \citenamefont
  {Kune{\v{s}}}}]{Ikeda.PRB.81.054502(2010)}%
  \BibitemOpen
  \bibfield  {author} {\bibinfo {author} {\bibfnamefont {H.}~\bibnamefont
  {Ikeda}}, \bibinfo {author} {\bibfnamefont {R.}~\bibnamefont {Arita}}, \ and\
  \bibinfo {author} {\bibfnamefont {J.}~\bibnamefont {Kune{\v{s}}}},\ }\href
  {\doibase 10.1103/PhysRevB.81.054502} {\bibfield  {journal} {\bibinfo
  {journal} {Phys. Rev. B}\ }\textbf {\bibinfo {volume} {81}},\ \bibinfo
  {pages} {054502} (\bibinfo {year} {2010})}\BibitemShut {NoStop}%
\bibitem [{\citenamefont {Charnukha}\ \emph {et~al.}(2015)\citenamefont
  {Charnukha}, \citenamefont {Evtushinsky}, \citenamefont {Matt}, \citenamefont
  {Xu}, \citenamefont {Shi}, \citenamefont {B\"uchner}, \citenamefont
  {Zhigadlo}, \citenamefont {Batlogg},\ and\ \citenamefont
  {Borisenko}}]{Charnukha2015}%
  \BibitemOpen
  \bibfield  {author} {\bibinfo {author} {\bibfnamefont {A.}~\bibnamefont
  {Charnukha}}, \bibinfo {author} {\bibfnamefont {D.}~\bibnamefont
  {Evtushinsky}}, \bibinfo {author} {\bibfnamefont {C.}~\bibnamefont {Matt}},
  \bibinfo {author} {\bibfnamefont {N.}~\bibnamefont {Xu}}, \bibinfo {author}
  {\bibfnamefont {M.}~\bibnamefont {Shi}}, \bibinfo {author} {\bibfnamefont
  {B.}~\bibnamefont {B\"uchner}}, \bibinfo {author} {\bibfnamefont {N.~D.}\
  \bibnamefont {Zhigadlo}}, \bibinfo {author} {\bibfnamefont {B.}~\bibnamefont
  {Batlogg}}, \ and\ \bibinfo {author} {\bibfnamefont {S.}~\bibnamefont
  {Borisenko}},\ }\href {\doibase 10.1038/srep18273} {\bibfield  {journal}
  {\bibinfo  {journal} {Sci. Rep.}\ }\textbf {\bibinfo {volume} {5}},\ \bibinfo
  {pages} {18273} (\bibinfo {year} {2015})}\BibitemShut {NoStop}%
\bibitem [{\citenamefont {Ortenzi}\ \emph {et~al.}(2009)\citenamefont
  {Ortenzi}, \citenamefont {Cappelluti}, \citenamefont {Benfatto},\ and\
  \citenamefont {Pietronero}}]{Ortenzi2009}%
  \BibitemOpen
  \bibfield  {author} {\bibinfo {author} {\bibfnamefont {L.}~\bibnamefont
  {Ortenzi}}, \bibinfo {author} {\bibfnamefont {E.}~\bibnamefont {Cappelluti}},
  \bibinfo {author} {\bibfnamefont {L.}~\bibnamefont {Benfatto}}, \ and\
  \bibinfo {author} {\bibfnamefont {L.}~\bibnamefont {Pietronero}},\ }\href
  {\doibase 10.1103/PhysRevLett.103.046404} {\bibfield  {journal} {\bibinfo
  {journal} {Phys. Rev. Lett.}\ }\textbf {\bibinfo {volume} {103}},\ \bibinfo
  {pages} {046404} (\bibinfo {year} {2009})}\BibitemShut {NoStop}%
\end{thebibliography}%


\section{Appendix: Tight-Binding Hamiltonian}
	
In this section we present the tight-binding Hamiltonian as obtained  with the $vasp2wannier$ package. This representation is applicable not only for the 1144 materials, but also for the 122 compounds, what underlines the similarity of both systems. In particular, the elements of Eq.(4) read

	\begin{align}
	\begin{aligned}
	H^\text{A1A1}_{11}&=t^{000}_{11}+2t^{010}_{11}\cos k_2+2t^{100}_{11}\cos k_1\\
	& \quad+2t^{020}_{11}\cos(2k_2)+2t^{200}_{11}\cos(2k_1)\\
	& \quad+4t^{110}_{11}\cos k_1\cos k_2\\
	H^\text{A1A1}_{12}&=4t^{110}_{12}\sin k_1\sin k_2\\
	H^\text{A1A1}_{13}&=2it^{010}_{13}\sin k_2-4it^{110}_{13}\cos k_1\sin k_2\\
	H^\text{A1A1}_{14}&=2it^{100}_{14}\sin k_1\\
	H^\text{A1A1}_{15}&=4t^{110}_{15}\sin k_1\sin k_2\\
	H^\text{A1A1}_{22}&=t^{000}_{22}+2t^{010}_{22}\cos k_2+2t^{100}_{22}\cos k_1\\
	& \quad+2t^{020}_{22}\cos(2k_2)+2t^{200}_{22}\cos(2k_1)\\
	& \quad+4t^{110}_{22}\cos k_1\cos k_2\\
	H^\text{A1A1}_{23}&=2it^{100}_{23}\sin k_1+2it^{200}_{23}\sin(2k_1)\\
	& \quad+4it^{110}_{23}\sin k_1\cos k_2\\
	& \quad+4it^{210}_{23}\sin(2k_1)\cos k_2\\
	H^\text{A1A1}_{24}&=2it^{010}_{24}\sin k_2+4it^{110}_{24}\cos k_1\sin k_2\\
	& \quad+4it^{120}_{24}\cos k_1\sin(2k_2)\\
	H^\text{A1A1}_{25}&=t^{000}_{25}+2t^{010}_{25}\cos k_2+2t^{100}_{25}\cos k_1\\
	& \quad+2t^{020}_{25}\cos(2k_2)+2t^{200}_{25}\cos(2k_1)\\
	& \quad+4t^{110}_{25}\cos k_1\cos k_2\\
	H^\text{A1A1}_{33}&=t^{000}_{33}+2t^{010}_{33}\cos k_2+2t^{100}_{33}\cos k_1\\
	& \quad+2t^{200}_{33}\cos(2k_1)+4t^{110}_{33}\cos k_1\cos k_2\\
	& \quad+2t^{300}_{33}\cos(3k_1)\\
	H^\text{A1A1}_{34}&=4t^{110}_{34}\sin k_1\sin k_2\\
	H^\text{A1A1}_{35}&=2it^{100}_{35}\sin k_1+2it^{200}_{35}\sin(2k_1)\\
	H^\text{A1A1}_{44}&=t^{000}_{44}+2t^{010}_{44}\cos k_2+2t^{100}_{44}\cos k_1\\
	& \quad+2t^{020}_{44}\cos(2k_2)+4t^{110}_{44}\cos k_1\cos k_2\\
	& \quad+2t^{030}_{44}\cos(3k_2)\\
	H^\text{A1A1}_{45}&=2it^{010}_{45}\sin k_2+2it^{020}_{45}\sin(2k_2)\\
	H^\text{A1A1}_{55}&=t^{000}_{55}+2t^{010}_{55}\cos k_2+2t^{100}_{55}\cos k_1\\
	& \quad+2t^{020}_{55}\cos(2k_2)+2t^{200}_{55}\cos(2k_1)\\
	& \quad+4t^{110}_{55}\cos k_1\cos k_2
		\end{aligned}
	\end{align}
and	
	\begin{align}
	\begin{aligned}
	H^\text{A1A2}_{16}&=2t^{000}_{16}(\cos k_x+\cos k_y)\\
	H^\text{A1A2}_{17}&=2t^{000}_{17}(\cos k_x-\cos k_y)\\
	& \quad+2t^{010}_{17}\left(\cos(2k_x-k_y)-\cos(k_x-2k_y)\right)\\
	& \quad+2t^{100}_{17}\left(-\cos(2k_x+k_y)+\cos(k_x+2k_y)\right)\\
	H^\text{A1A2}_{18}&=2it^{000}_{18}(\sin k_x-\sin k_y)\\
	& \quad+2it^{100}_{18}(\sin(2k_x+k_y)-\sin(k_x+2k_y))\\
	& \quad+2it^{200}_{18}(\sin(3k_x+2k_y)-\sin(2k_x+3k_y))\\
	H^\text{A1A2}_{19}&=2it^{000}_{19}(\sin k_x+\sin k_y)\\
	& \quad+2it^{010}_{19}(\sin(2k_x-k_y)-\sin(k_x-2k_y))\\
	& \quad+2it^{020}_{19}(\sin(3k_x-2k_y)-\sin(2k_x-3k_y))\\
	H^\text{A1A2}_{1,10}&=2t^{000}_{1,10}(-\cos k_x+\cos k_y)\\
	& \quad+2t^{010}_{1,10}\left(-\cos(2k_x-k_y)+\cos(k_x-2k_y)\right)\\
	& \quad+2t^{100}_{1,10}(-\cos(2k_x+k_y)+\cos(k_x+2k_y))\\
	H^\text{A1A2}_{27}&=-2t^{000}_{27}(\cos k_x+\cos k_y)\\
	& \quad+4t^{010}_{27}(\cos(2k_x)\cos k_y+\cos k_x\cos(2k_y))\\
	H^\text{A1A2}_{28}&=2it^{000}_{28}(\sin k_x+\sin k_y)\\
	& \quad+2it^{010}_{28}\left(\sin(2k_x-k_y)-\sin(k_x-2k_y)\right)\\
	& \quad+2it^{100}_{28}\left(\sin(2k_x+k_y)+\sin(k_x+2k_y)\right)\\
	H^\text{A1A2}_{29}&=2it^{000}_{29}(-\sin k_x+\sin k_y)\\
	& \quad-2it^{010}_{29}(\sin(2k_x-k_y)+\sin(k_x-2k_y))\\
	& \quad+2it^{100}_{29}\left(-\sin(2k_x+k_y)+\sin(k_x+2k_y)\right)\\
	H^\text{A1A2}_{2,10}&=-2t^{000}_{2,10}(\cos k_x+\cos k_y)\\
	& \quad-2t^{010}_{2,10}(\cos(2k_x-k_y)+\cos(k_x-2k_y))\\
	& \quad+2t^{100}_{2,10}(\cos(2k_x+k_y)+\cos(k_x+2k_y))\\
	H^\text{A1A2}_{38}&=-2t^{000}_{38}(\cos k_x+\cos k_y)\\
	& \quad+2t^{010}_{38}(\cos(2k_x-k_y)+\cos(k_x-2k_y))\\
	& \quad-2t^{100}_{38}(\cos(2k_x+k_y)+\cos(k_x+2k_y))\\
	H^\text{A1A2}_{39}&=2t^{000}_{39}(-\cos k_x+\cos k_y)\\
	& \quad+4t^{010}_{39}(\cos(2k_x)\cos k_y-\cos k_x\cos(2k_y))\\
	H^\text{A1A2}_{3,10}&=2it^{000}_{3,10}(\sin k_x+\sin k_y)\\
	H^\text{A1A2}_{48}&=2t^{000}_{48}(-\cos k_x+\cos k_y)\\
	& \quad+4t^{010}_{48}(\cos(2k_x)\cos k_y-\cos k_x\cos(2k_y))\\
	H^\text{A1A2}_{49}&=-2t^{000}_{49}(\cos k_x+\cos k_y)\\
	& \quad-2t^{010}_{49}(\cos(2k_x-k_y)+\cos(k_x-2k_y))\\
	& \quad+2t^{100}_{49}(\cos(2k_x+k_y)+\cos(k_x+2k_y))\\
	H^\text{A1A2}_{4,10}&=2it^{000}_{4,10}(-\sin k_x+\sin k_y)\\
	& \quad-2it^{010}_{4,10}(\sin(2k_x-k_y)+\sin (k_x-2k_y))\\
	H^\text{A1A2}_{5,10}&=2t^{000}_{5,10}(\cos k_x+\cos k_y)\\
	& \quad-4t^{010}_{5,10}(\cos(2k_x)\cos k_y+\cos k_x\cos(2k_y))\\
	& \quad+2t^{110}_{5,10}(\cos(3k_x)+\cos(3k_y)) \quad.
	\end{aligned}
	\end{align}
		\begin{align}
		\begin{aligned}
		H^\text{A1B1}_{1,11}&=t^{000}_{1,11}+2t^{010}_{1,11}\cos k_2+2t^{100}_{1,11}\cos k_1\\
		& \quad+4t^{110}_{1,11}\cos k_1\cos k_2\\
		H^\text{A1B1}_{1,12}&=0\\
		H^\text{A1B1}_{1,13}&=2it^{010}_{1,13}\sin k_2\\
		H^\text{A1B1}_{1,14}&=0\\
		H^\text{A1B1}_{1,15}&=0\\
		H^\text{A1B1}_{2,12}&=t^{000}_{2,12}+2t^{010}_{2,12}\cos k_2+2t^{100}_{2,12}\cos k_1\\
		& \quad+2t^{020}_{2,12}\cos(2k_2)\\
		H^\text{A1B1}_{2,13}&=2it^{100}_{2,13}\sin k_1\\
		H^\text{A1B1}_{2,14}&=2it^{010}_{2,14}\sin k_2+4it^{110}_{2,14}\cos k_1\sin k_2\\
		H^\text{A1B1}_{2,15}&=t^{000}_{2,15}+2t^{010}_{2,15}\cos k_2+2t^{100}_{2,15}\cos k_1\\
		& \quad+2t^{020}_{2,15}\cos(2k_2)\\
		H^\text{A1B1}_{3,13}&=t^{000}_{3,13}+2t^{010}_{3,13}\cos k_2\\
		& \quad+4t^{110}_{3,13}\cos k_1\cos k_2\\
		H^\text{A1B1}_{3,14}&=0\\
		H^\text{A1B1}_{3,15}&=2it^{100}_{3,15}\sin k_1\\
		H^\text{A1B1}_{4,14}&=t^{000}_{4,14}+2t^{010}_{4,14}\cos k_2+2t^{100}_{4,14}\cos k_1\\
		& \quad+4t^{110}_{4,14}\cos k_1\cos k_2\\
		& \quad+4t^{120}_{4,14}\cos k_1\cos(2k_2)\\
		H^\text{A1B1}_{4,15}&=4it^{110}_{4,15}\cos k_1\sin k_2\\
		H^\text{A1B1}_{5,15}&=t^{000}_{5,15}+2t^{010}_{5,15}\cos k_2+2t^{100}_{5,15}\cos k_1\\
		& \quad+2t^{020}_{5,15}\cos(2k_2)+4t^{110}_{5,15}\cos k_1\cos k_2
		\end{aligned}
		\end{align}
and for the A1B2 block
		\begin{align}
		\begin{aligned}
		H^\text{A1B2}_{1,16}&=0\\
		H^\text{A1B2}_{1,17}&=2t^{000}_{1,17}(\cos k_x-\cos k_y)\\
		H^\text{A1B2}_{1,18}&=2it^{000}_{1,18}(\sin k_x-\sin k_y)\\
		& \quad-2it^{010}_{1,18}(\sin(2k_x-k_y)+\sin(k_x-2k_y))\\
		& \quad+2it^{100}_{1,18}(\sin(2k_x+k_y)-\sin(k_x+2k_y))\\
		H^\text{A1B2}_{1,19}&=0\\
		H^\text{A1B2}_{1,20}&=0\\
		H^\text{A1B2}_{2,17}&=2t^{000}_{2,17}(\cos k_x+\cos k_y)\\
		H^\text{A1B2}_{2,18}&=-2it^{000}_{2,18}(\sin k_x+\sin k_y)\\
		& \quad+2it^{010}_{2,18}(\sin(2k_x-k_y)-\sin(k_x-2k_y))\\
		H^\text{A1B2}_{2,19}&=2it^{000}_{2,19}(-\sin k_x+\sin k_y)\\
		& \quad+2it^{010}_{2,19}(\sin(2k_x-k_y)+\sin(k_x-2k_y))\\
		H^\text{A1B2}_{2,20}&=2t^{000}_{2,20}(\cos k_x+\cos k_y)\\
		& \quad+2t^{010}_{2,20}(\cos(2k_x-k_y)+\cos(k_x-2k_y))\\
		H^\text{A1B2}_{3,18}&=-2t^{000}_{3,18}(\cos k_x+\cos k_y)\\
		& \quad+2t^{110}_{3,18}(\cos(3k_x)+\cos(3k_y))\\
		H^\text{A1B2}_{3,19}&=0\\
		H^\text{A1B2}_{3,20}&=-2it^{000}_{3,20}(\sin k_x+\sin k_y)\\
		H^\text{A1B2}_{4,19}&=H^\text{A1B2}_{3,18}\\
		H^\text{A1B2}_{4,20}&=2it^{000}_{4,20}(\sin k_x-\sin k_y)\\
		H^\text{A1B2}_{5,20}&=-2t^{000}_{5,20}(\cos k_x+\cos k_y)\\
		& \quad+4t^{010}_{5,20}(\cos(2k_x)\cos k_y+\cos k_x\cos(2k_y))
		\end{aligned}
		\end{align}
The dispersion part involving $k_z$ direction has for A1A1 term
		\begin{align}
		H^\text{A1A1}_{22}=2t^{001}_{22}\cos k_z
		\end{align}
and A1B1 term
		\begin{align}
		\begin{aligned}
		H^\text{A1B1}_{1,11}&=+\big(t^{001}_{1,11}+2t^{011}_{1,11}\cos k_2+2t^{101}_{1,11}\cos k_1\\
		& \quad+4t^{111}_{1,11}\cos k_1\cos k_2\big)e^{-ik_z}\\
		H^\text{A1B1}_{1,12}&=0\\
		H^\text{A1B1}_{1,13}&=0\\
		H^\text{A1B1}_{1,14}&=+2it^{101}_{1,14}\sin k_1\, e^{-ik_z}\\
		H^\text{A1B1}_{1,15}&=0\\
		H^\text{A1B1}_{2,12}&=+\big(t^{001}_{2,12}+2t^{011}_{2,12}\cos k_2+2t^{101}_{2,12}\cos k_1\\
		& \quad+2t^{201}_{2,12}\cos(2k_1)\\
		& \quad+4t^{111}_{2,12}\cos k_1\cos k_2\big)e^{-ik_z}\\
		H^\text{A1B1}_{2,13}&=+2it^{101}_{2,13}\sin k_1\, e^{-ik_z}\\
		H^\text{A1B1}_{2,14}&=0\\
		H^\text{A1B1}_{2,15}&=+\big(t^{001}_{2,15}+2t^{011}_{2,15}\cos k_2+2t^{101}_{2,15}\cos k_1\\
		& \quad+2t^{201}_{2,15}\cos(2k_1)\big)e^{-ik_z}\\
		H^\text{A1B1}_{3,13}&=+\big(t^{001}_{3,13}+2t^{011}_{3,13}\cos k_2+2t^{101}_{3,13}\cos k_1\\
		& \quad+4t^{111}_{3,13}\cos k_1\cos k_2\\
		& \quad+4t^{211}_{3,13}\cos(2k_1)\cos k_2\big)e^{-ik_z}\\
		H^\text{A1B1}_{3,14}&=0\\
		H^\text{A1B1}_{3,15}&=+\big(2it^{101}_{3,15}\sin k_1-4it^{111}_{3,15}\sin k_1\cos k_2\big)e^{-ik_z}\\
		H^\text{A1B1}_{4,14}&=+\big(t^{001}_{4,14}+2t^{011}_{4,14}\cos k_2+2t^{101}_{4,14}\cos k_1\\
		& \quad+4t^{111}_{4,14}\cos k_1\cos k_2\big)e^{-ik_z}\\
		H^\text{A1B1}_{4,15}&=+\big(2it^{011}_{4,15}\sin k_2+4it^{111}_{4,15}\cos k_1\sin k_2\big)e^{-ik_z}\\
		H^\text{A1B1}_{5,15}&=+\big(t^{001}_{5,15}+2t^{011}_{5,15}\cos k_2+2t^{101}_{5,15}\cos k_1\\
		& \quad+2t^{201}_{5,15}\cos(2k_1)\\
		& \quad+4t^{111}_{5,15}\cos k_1\cos k_2\big)e^{-ik_z}
		\end{aligned}
		\end{align}
as well as A1B2 part
		\begin{align}
		\begin{aligned}
		H^\text{A1B2}_{1,16}&=0\\
		H^\text{A1B2}_{1,17}&=+2t^{001}_{1,17}(-\cos k_x+\cos k_y)e^{-ik_z}\\
		H^\text{A1B2}_{1,18}&=0\\
		H^\text{A1B2}_{1,19}&=+\big[2it^{001}_{1,19}(\sin k_x+\sin k_y)\\
		& \quad+2it^{011}_{1,19}(\sin(2k_x-k_y)-\sin(k_x-2k_y))\\
		& \quad-2it^{101}_{1,19}(\sin(2k_x+k_y)\\
		& \quad+\sin(k_x+2k_y))\big]e^{-ik_z}\\
		H^\text{A1B2}_{1,20}&=+2t^{001}_{1,20}(-\cos k_x+\cos k_y)e^{-ik_z}\\
		H^\text{A1B2}_{2,17}&=+2t^{001}_{2,17}(\cos k_x+\cos k_y)e^{-ik_z}\\
		H^\text{A1B2}_{2,18}&=+\big[2it^{001}_{2,18}(\sin k_x+\sin k_y)\\
		& \quad-2it^{101}_{2,18}(\sin(2k_x+k_y)\\
		& \quad+\sin(k_x+2k_y))\big]e^{-ik_z}\\
		H^\text{A1B2}_{2,19}&=+\big[2it^{001}_{2,19}(\sin k_x-\sin k_y)\\
		& \quad+2it^{101}_{2,19}(-\sin(2k_x+k_y)\\
		& \quad+\sin(k_x+2k_y))\big]e^{-ik_z}\\
		H^\text{A1B2}_{2,20}&=+\big[2t^{001}_{2,20}(\cos k_x+\cos k_y)\\
		& \quad-2t^{101}_{2,20}(\cos(2k_x+k_y)\\
		& \quad+\cos(k_x+2k_y))\big]e^{-ik_z}\\
		H^\text{A1B2}_{3,18}&=+\big[-2t^{001}_{3,18}(\cos k_x+\cos k_y)\\
		& \quad+2t^{111}_{3,18}(\cos(3k_x)+\cos(3k_y))\big]e^{-ik_z}\\
		H^\text{A1B2}_{3,19}&=0\\
		H^\text{A1B2}_{3,20}&=+\big[2it^{001}_{3,20}(\sin k_x+\sin k_y)\\
		& \quad+2it^{011}_{3,20}(-\sin(2k_x-k_y)\\
		& \quad+\sin(k_x-2k_y))\big]e^{-ik_z}\\
		H^\text{A1B2}_{4,19}&=H^\text{A1B2}_{3,18}\\
		H^\text{A1B2}_{4,20}&=+\big[2it^{001}_{4,20}(-\sin k_x+\sin k_y)\\
		& \quad-2it^{011}_{4,20}(\sin(2k_x-k_y)\\
		& \quad+\sin(k_x-2k_y))\big]e^{-ik_z}\\
		H^\text{A1B2}_{5,20}&=+\big[-2t^{001}_{5,20}(\cos k_x+\cos k_y)\\
		& \quad+4t^{101}_{5,20}(\cos(2k_x)\cos k_y\\
		& \quad+\cos k_x\cos(2k_y))\big]e^{-ik_z}
		\end{aligned}
		\end{align}
		
Below we give the parameters of the hopping integrals for \CaK \, system for the	A1A1 term
		\begin{align}
		\begin{aligned}
		t^{000}_{11}&=0.247088,& t^{010}_{11}&=-0.128852,\\
		t^{100}_{11}&=-0.015027,& t^{020}_{11}&=0.019461,\\
		t^{200}_{11}&=0.022313,& t^{110}_{11}&=-0.036534,\\
		t^{110}_{12}&=0.018133,& t^{010}_{13}&=0.174436,\\
		t^{110}_{13}&=0.010374,& t^{100}_{14}&=-0.126588,\\
		t^{110}_{15}&=0.019849,& t^{000}_{22}&=0.040038,\\
		t^{010}_{22}&=0.233719,& t^{100}_{22}&=-0.084639,\\
		t^{020}_{22}&=-0.041893,& t^{200}_{22}&=0,\\
		t^{110}_{22}&=-0.017867,& t^{100}_{23}&=-0.058680,\\
		t^{200}_{23}&=-0.028825,& t^{110}_{23}&=0,\\
		t^{210}_{23}&=0,& t^{010}_{24}&=0.141905,\\
		t^{110}_{24}&=0,& t^{120}_{24}&=0,\\
		t^{000}_{25}&=-0.212933,& t^{010}_{25}&=-0.051335,\\
		t^{100}_{25}&=-0.085047,& t^{020}_{25}&=0.013004,\\
		t^{200}_{25}&=0.016189,& t^{110}_{25}&=-0.013983,\\
		t^{000}_{33}&=0.183023,& t^{010}_{33}&=0.140469,\\
		t^{100}_{33}&=0.336148& t^{110}_{33}&=-0.015620,\\
		t^{200}_{33}&=0.077006,& t^{300}_{33}&=0.022415,\\
		t^{110}_{34}&=-0.032167,& t^{100}_{35}&=0.201822,\\
		t^{200}_{35}&=0,& t^{000}_{44}&=0.147106,\\
		t^{010}_{44}&=0.429681,& t^{100}_{44}&=0.142601,\\
		t^{110}_{44}&=-0.011815,& t^{020}_{44}&=0.082600,\\
		t^{030}_{44}&=0.021224,& t^{010}_{45}&=-0.078445,\\
		t^{020}_{45}&=-0.024122,& t^{000}_{55}&=0.067747,\\
		t^{010}_{55}&=-0.112947,& t^{100}_{55}&=0.248170,\\
		t^{110}_{55}&=-0.024991,& t^{200}_{55}&=-0.043234,\\
		t^{020}_{55}&=0
		\end{aligned}\label{Eq:Ca1144A1A1}
		\end{align}
and  A1A2
		\begin{align}
		\begin{aligned}
		t^{000}_{16}&=-0.381198,& t^{000}_{17}&=0.202289,\\
		t^{010}_{17}&=0.018080,& t^{100}_{17}&=0,\\
		t^{000}_{18}&=0.347157,& t^{100}_{18}&=0.039476,\\
		t^{200}_{18}&=0.011114,& t^{000}_{19}&=0.254869,\\
		t^{010}_{19}&=0.022287,& t^{020}_{19}&=0.010527,\\
		t^{000}_{1,10}&=0.264651,& t^{010}_{1,10}&=0,\\
		t^{100}_{1,10}&=0.017994,& t^{000}_{27}&=0.111094,\\
		t^{010}_{27}&=0.027008,& t^{000}_{28}&=0.204372,\\
		t^{010}_{28}&=0,& t^{100}_{28}&=0,\\
		t^{000}_{29}&=0.081707,& t^{010}_{29}&=0.021110,\\
		t^{100}_{29}&=0,& t^{000}_{2,10}&=0.066480,\\
		t^{010}_{2,10}&=0,& t^{100}_{2,10}&=0.028756,\\
		t^{000}_{38}&=0.229540,& t^{010}_{38}&=0,\\
		t^{100}_{38}&=0.035261,& t^{000}_{39}&=0.103442,\\
		t^{010}_{39}&=0.032044,& t^{000}_{3,10}&=0.220496,\\
		t^{000}_{48}&=0.167758,& t^{010}_{48}&=0.019836,\\
		t^{000}_{49}&=t^{000}_{38},& t^{010}_{49}&=t^{100}_{38},\\
		t^{100}_{49}&=t^{010}_{38},& t^{000}_{4,10}&=0.054671,\\
		t^{010}_{4,10}&=0.021606,& t^{000}_{5,10}&=0.113811,\\
		t^{010}_{5,10}&=0.029265,& t^{110}_{5,10}&=0
		\end{aligned}
		\end{align}
		The parameters for A1B1 block read
		\begin{align}
		\begin{aligned}
		t^{000}_{1,11}&=-0.014187,& t^{010}_{1,11}&=0,\\
		t^{100}_{1,11}&=0,& t^{110}_{1,11}&=0,\\
		t^{010}_{1,13}&=0,& t^{000}_{2,12}&=-0.178573,\\
		t^{010}_{2,12}&=-0.079501,& t^{100}_{2,12}&=0,\\
		t^{020}_{2,12}&=0,& t^{100}_{2,13}&=-0.011803,\\
		t^{010}_{2,14}&=-0.012294,& t^{110}_{2,14}&=0.010139,\\
		t^{000}_{2,15}&=0.022342,& t^{010}_{2,15}&=0,\\
		t^{100}_{2,15}&=0.017056,& t^{020}_{2,15}&=0,\\
		t^{000}_{3,13}&=0.018604,& t^{010}_{3,13}&=0.010938,\\
		t^{110}_{3,13}&=0,& t^{100}_{3,15}&=0,\\
		t^{000}_{4,14}&=0.022463,& t^{010}_{4,14}&=0,\\
		t^{100}_{4,14}&=0,& t^{110}_{4,14}&=0,\\
		t^{120}_{4,14}&=0,& t^{110}_{4,15}&=0,\\
		t^{000}_{5,15}&=-0.015017,& t^{010}_{5,15}&=0,\\
		t^{100}_{5,15}&=-0.011221,& t^{110}_{5,15}&=0,\\
		t^{020}_{5,15}&=0
		\end{aligned}
		\end{align}
		and  A1B2
		\begin{align}
		\begin{aligned}
		t^{000}_{1,17}&=0,& t^{000}_{1,18}&=0,\\
		t^{010}_{1,18}&=0,& t^{100}_{1,18}&=0,\\
		t^{000}_{2,17}&=0.084500,& t^{000}_{2,18}&=0,\\
		t^{010}_{2,18}&=0.010832,& t^{000}_{2,19}&=0,\\
		t^{010}_{2,19}&=0,& t^{000}_{2,20}&=0.015346,\\
		t^{010}_{2,20}&=0,& t^{000}_{3,18}&=0,\\
		t^{110}_{3,18}&=0,& t^{000}_{3,20}&=0,\\
		t^{000}_{4,20}&=0,& t^{000}_{5,20}&=0,\\
		t^{010}_{5,20}&=0
		\end{aligned}
		\end{align}
Similarly, we find parameters A1A1 3d
		\begin{align}
		t^{001}_{22}&=0
		\end{align}
		A1B1 3d
		\begin{align}
		\begin{aligned}
		t^{001}_{1,11}&=-0.061912,& t^{011}_{1,11}&=-0.014910,\\
		t^{101}_{1,11}&=0.032455,& t^{111}_{1,11}&=0.013021,\\
		t^{101}_{1,14}&=-0.033288,& t^{001}_{2,12}&=-0.040849,\\
		t^{011}_{2,12}&=-0.018842,& t^{101}_{2,12}&=0.021234,\\
		t^{111}_{2,12}&=0.010619,& t^{201}_{2,12}&=0,\\
		t^{101}_{2,13}&=-0.028809,& t^{001}_{2,15}&=0.055394,\\
		t^{011}_{2,15}&=0.028108,& t^{101}_{2,15}&=0.012610,\\
		t^{201}_{2,15}&=0,& t^{001}_{3,13}&=0.117064,\\
		t^{011}_{3,13}&=-0.014458,& t^{101}_{3,13}&=0.063591,\\
		t^{111}_{3,13}&=-0.017604,& t^{211}_{3,13}&=-0.017878,\\
		t^{101}_{3,15}&=0.027871,& t^{111}_{3,15}&=0.015766,\\
		t^{001}_{4,14}&=0.077263,& t^{011}_{4,14}&=-0.018486,\\
		t^{101}_{4,14}&=0.034123,& t^{111}_{4,14}&=0,\\
		t^{011}_{4,15}&=0.029779,& t^{111}_{4,15}&=0.011809,\\
		t^{001}_{5,15}&=-0.354701,& t^{011}_{5,15}&=0.021342,\\
		t^{101}_{5,15}&=-0.164139,& t^{111}_{5,15}&=0,\\
		t^{201}_{5,15}&=0.014937
		\end{aligned}
		\end{align}
	and parameters A1B2 3d
		\begin{align}
		\begin{aligned}
		t^{001}_{1,17}&=0.016345,& t^{001}_{1,19}&=0.049462,\\
		t^{011}_{1,19}&=0.012258,& t^{101}_{1,19}&=0,\\
		t^{001}_{1,20}&=0.010052,& t^{001}_{2,17}&=0.012790,\\
		t^{001}_{2,18}&=0.027296,& t^{101}_{2,18}&=0,\\
		t^{001}_{2,19}&=0,& t^{101}_{2,19}&=0,\\
		t^{001}_{2,20}&=0.028421,& t^{101}_{2,20}&=0,\\
		t^{001}_{3,18}&=0.044275,& t^{111}_{3,18}&=0.010797,\\
		t^{001}_{3,20}&=0,& t^{011}_{3,20}&=0.011906,\\
		t^{001}_{4,20}&=0,& t^{011}_{4,20}&=0.013628,\\
		t^{001}_{5,20}&=0.167859,& t^{101}_{5,20}&=0
		\end{aligned}
		\end{align}
		
The same parameters for \Ca \, are
		
		Parameter A1A1
		\begin{align}
		\begin{aligned}
		t^{000}_{11}&=-0.365859,& t^{010}_{11}&=-0.052564,\\
		t^{100}_{11}&=t^{010}_{11},& t^{110}_{11}&=-0.019398,\\
		t^{020}_{11}&=0.016116,& t^{200}_{11}&=t^{020}_{11},\\
		t^{110}_{12}&=0,& t^{010}_{13}&=0.120333,\\
		t^{110}_{13}&=0,& t^{100}_{14}&=-t^{010}_{13},\\
		t^{110}_{15}&=0.019057,& t^{000}_{22}&=0.264452,\\
		t^{010}_{22}&=0.161138,& t^{100}_{22}&=t^{010}_{22},\\
		t^{110}_{22}&=0.013436,& t^{020}_{22}&=-0.040903,\\
		t^{200}_{22}&=t^{020}_{22},& t^{100}_{23}&=0.068608,\\
		t^{110}_{23}&=0.013523,& t^{210}_{23}&=-0.011030,\\
		t^{200}_{23}&=0,& t^{010}_{24}&=t^{100}_{23},\\
		t^{110}_{24}&=t^{110}_{23},& t^{120}_{24}&=t^{210}_{23},\\
		t^{000}_{25}&=0,& t^{010}_{25}&=0.174989,\\
		t^{100}_{25}&=-t^{010}_{25},& t^{110}_{25}&=0,\\
		t^{020}_{25}&=-0.020306,& t^{200}_{25}&=-t^{020}_{25},\\
		t^{000}_{33}&=0.012164,& t^{010}_{33}&=0.104970,\\
		t^{100}_{33}&=0.324250,& t^{110}_{33}&=0,\\
		t^{200}_{33}&=0.069523,& t^{300}_{33}&=0.017765,\\
		t^{110}_{34}&=-0.038357,& t^{100}_{35}&=0.151463,\\
		t^{200}_{35}&=0.031583,& t^{000}_{44}&=t^{000}_{33},\\
		t^{010}_{44}&=t^{100}_{33},& t^{100}_{44}&=t^{010}_{33},\\
		t^{110}_{44}&=0,& t^{020}_{44}&=t^{200}_{33},\\
		t^{030}_{44}&=t^{300}_{33},& t^{010}_{45}&=-t^{100}_{35},\\
		t^{020}_{45}&=-t^{200}_{35},& t^{000}_{55}&=-0.089032,\\
		t^{010}_{55}&=0,& t^{100}_{55}&=0,\\
		t^{110}_{55}&=-0.033271,& t^{020}_{55}&=-0.015254,\\
		t^{200}_{55}&=t^{020}_{55}
		\end{aligned} \label{Eq:Ca122A1A1}
		\end{align}
		
		Parameter A1A2
		\begin{align}
		\begin{aligned}
		t^{000}_{16}&=-0.363661,& t^{000}_{17}&=0,\\
		t^{010}_{17}&=0.010562,& t^{100}_{17}&=t^{010}_{17},\\
		t^{000}_{18}&=0.244647,& t^{100}_{18}&=0.022411,\\
		t^{200}_{18}&=0,& t^{000}_{19}&=t^{000}_{18},\\
		t^{010}_{19}&=t^{100}_{18},& t^{020}_{19}&=t^{200}_{18},\\
		t^{000}_{1,10}&=0.304047,& t^{010}_{1,10}&=0.011386,\\
		t^{100}_{1,10}&=t^{010}_{1,10},& t^{000}_{27}&=0.211970,\\
		t^{010}_{27}&=0.035072,& t^{000}_{28}&=0.154401,\\
		t^{010}_{28}&=0.014769,& t^{100}_{28}&=0.021142,\\
		t^{000}_{29}&=t^{000}_{28},& t^{010}_{29}&=t^{100}_{28},\\
		t^{100}_{29}&=t^{010}_{28},& t^{000}_{2,10}&=0,\\
		t^{010}_{2,10}&=0.019052,& t^{100}_{2,10}&=t^{010}_{2,10},\\
		t^{000}_{38}&=0.183874,& t^{010}_{38}&=0.011643,\\
		t^{100}_{38}&=0.041421,& t^{000}_{39}&=0.113111,\\
		t^{010}_{39}&=0.023085,& t^{000}_{3,10}&=0.092244,\\
		t^{000}_{48}&=t^{000}_{39},& t^{010}_{48}&=t^{010}_{39},\\
		t^{000}_{49}&=t^{000}_{38},& t^{010}_{49}&=t^{100}_{38},\\
		t^{100}_{49}&=t^{010}_{38},& t^{000}_{4,10}&=-t^{000}_{3,10},\\
		t^{010}_{4,10}&=0,& t^{000}_{5,10}&=0.059586,\\
		t^{010}_{5,10}&=0.024878,& t^{110}_{5,10}&=0.013588
		\end{aligned}
		\end{align}
		
		Parameter A1B1
		\begin{align}
		\begin{aligned}
		t^{000}_{1,11}&=-0.062812,& t^{010}_{1,11}&=0.035863,\\
		t^{100}_{1,11}&=-0.018856,& t^{110}_{11}&=0.016659,\\
		t^{010}_{1,13}&=0.030549,& t^{000}_{2,12}&=-0.262053,\\
		t^{010}_{2,12}&=-0.072906,& t^{100}_{2,12}&=-0.025107,\\
		t^{020}_{2,12}&=0.010635,& t^{100}_{2,13}&=0,\\
		t^{010}_{2,14}&=-0.056892,& t^{110}_{2,14}&=0,\\
		t^{000}_{2,15}&=-0.170028,& t^{010}_{2,15}&=-0.090345,\\
		t^{100}_{2,15}&=0.031744,& t^{020}_{2,15}&=0.012453,\\
		t^{000}_{3,13}&=0.068646,& t^{010}_{3,13}&=0.045685,\\
		t^{110}_{3,13}&=-0.011647,& t^{100}_{3,15}&=-0.018612,\\
		t^{000}_{4,14}&=0.110209,& t^{010}_{4,14}&=0.059951,\\
		t^{100}_{4,14}&=-0.016171,& t^{110}_{4,14}&=-0.018985,\\
		t^{120}_{4,14}&=-0.016778,& t^{110}_{4,15}&=0.020444,\\
		t^{000}_{5,15}&=-0.156738,& t^{010}_{5,15}&=-0.062072,\\
		t^{100}_{5,15}&=0.037088,& t^{110}_{5,15}&=0.021684,\\
		t^{020}_{5,15}&=0.010814
		\end{aligned}
		\end{align}
		
		Parameter A1B2
		\begin{align}
		\begin{aligned}
		t^{000}_{1,17}&=0.020947,& t^{000}_{1,18}&=0.051860,\\
		t^{010}_{1,18}&=0.011344,& t^{100}_{1,18}&=0.012005,\\
		t^{000}_{2,17}&=0.119800,& t^{000}_{2,18}&=0.022801,\\
		t^{010}_{2,18}&=0.014765,& t^{000}_{2,19}&=0.030462,\\
		t^{010}_{2,19}&=0.012601,& t^{000}_{2,20}&=-0.083382,\\
		t^{010}_{2,20}&=0.011417,& t^{000}_{3,18}&=0.045360,\\
		t^{110}_{3,18}&=0.011536,& t^{000}_{3,20}&=0.023115,\\
		t^{000}_{4,20}&=0.011132,& t^{000}_{5,20}&=0.072867,\\
		t^{010}_{5,20}&=0.010693
		\end{aligned}
		\end{align}
		
		Parameter A1A1 3d
		\begin{align}
		t^{001}_{22}&=0.016488
		\end{align}
		
		Parameter A1B1 3d
		\begin{align}
		\begin{aligned}
		t^{001}_{1,11}&=t^{000}_{1,11},& t^{011}_{1,11}&=t^{100}_{1,11},\\
		t^{101}_{1,11}&=t^{010}_{1,11},& t^{111}_{1,11}&=t^{110}_{1,11},\\
		t^{101}_{1,14}&=-t^{010}_{1,13},& t^{001}_{2,12}&=t^{000}_{2,12},\\
		t^{011}_{2,12}&=t^{100}_{2,12},& t^{101}_{2,12}&=t^{010}_{2,12},\\
		t^{111}_{2,12}&=0,& t^{201}_{2,12}&=t^{020}_{2,12},\\
		t^{101}_{2,13}&=t^{010}_{2,14},& t^{001}_{2,15}&=-t^{000}_{2,15},\\
		t^{011}_{2,15}&=-t^{100}_{2,15},& t^{101}_{2,15}&=-t^{010}_{2,15},\\
		t^{201}_{2,15}&=-t^{020}_{2,15},& t^{001}_{3,13}&=t^{000}_{4,14},\\
		t^{011}_{3,13}&=t^{100}_{4,14},& t^{101}_{3,13}&=t^{010}_{4,14},\\
		t^{111}_{3,13}&=t^{110}_{4,14},& t^{211}_{3,13}&=t^{120}_{4,14},\\
		t^{101}_{3,15}&=0,& t^{110}_{3,15}&=t^{110}_{4,15},\\
		t^{001}_{4,14}&=t^{000}_{3,13},& t^{011}_{4,14}&=0,\\
		t^{101}_{4,14}&=t^{010}_{3,13},& t^{111}_{4,14}&=t^{110}_{3,13},\\
		t^{011}_{4,15}&=-t^{100}_{3,15},& t^{111}_{4,15}&=0,\\
		t^{001}_{5,15}&=t^{000}_{5,15},& t^{011}_{5,15}&=t^{100}_{5,15},\\
		t^{101}_{5,15}&=t^{010}_{5,15},& t^{111}_{5,15}&=t^{110}_{5,15},\\
		t^{201}_{5,15}&=t^{020}_{5,15}
		\end{aligned}
		\end{align}\\
		
		Parameter A1B2 3d
		\begin{align}
		\begin{aligned}
		t^{001}_{1,17}&=t^{000}_{1,17},& t^{001}_{1,19}&=t^{000}_{1,18},\\
		t^{011}_{1,19}&=t^{100}_{1,18},& t^{101}_{1,19}&=t^{010}_{1,18},\\
		t^{001}_{1,20}&=0,& t^{001}_{2,17}&=t^{000}_{2,17},\\
		t^{001}_{2,18}&=t^{000}_{2,19},& t^{101}_{2,18}&=t^{010}_{2,19},\\
		t^{001}_{2,19}&=t^{000}_{2,18},& t^{101}_{2,19}&=t^{010}_{2,18},\\
		t^{001}_{2,20}&=-t^{000}_{2,20},& t^{101}_{2,20}&=t^{010}_{2,20},\\
		t^{001}_{3,18}&=t^{000}_{3,18},& t^{111}_{3,18}&=t^{110}_{3,18},\\
		t^{001}_{3,20}&=t^{000}_{3,20},& t^{011}_{3,20}&=0,\\
		t^{001}_{4,20}&=t^{000}_{3,20},& t^{011}_{4,20}&=0,\\
		t^{001}_{5,20}&=t^{000}_{5,20},& t^{101}_{5,20}&=t^{010}_{5,20}
		\end{aligned}
		\end{align}

\end{document}